\documentclass{article}
\usepackage{amsfonts,amssymb,amsthm}

\newcommand{\R}{\mathbb{R}}
\newcommand{\C}{\mathbb{C}}
\newcommand{\I}{\mathrm{i}}
\newcommand{\E}{\mathrm{e}}
\newcommand{\D}{\mathrm{d}}
\newcommand{\N}{\mathbb{N}}

\newcommand{\T}{\mathbb{T}}
\newcommand{\Hi}{\mathcal{H}}
\newcommand{\B}{\mathcal{B}}
\newcommand{\M}{\mathcal{M}}
\newcommand{\U}{\mathcal{U}}
\newcommand{\F}{\mathcal{F}}
\newcommand{\K}{\mathcal{K}}

\newcommand{\Do}{\mathcal{D}}
\newcommand{\Sch}{\mathcal{S}}
\newcommand{\A}{\mathcal{A}}
\newcommand{\Or}{\mathcal{O}}
\newcommand{\Lap}{\Delta}
\newcommand{\epsi}{\varepsilon}
\newcommand{\ph}{\varphi}
\newcommand{\ix}{\index}
\newcommand{\indexd}{\in\{ 1,\ldots,d \}}
\newcommand{\Ht}{\mathcal{H}_{\tau}}
\newcommand{\vp}{\times}

\newcommand{\Href}{\mathcal{H}_{\mathrm{ref}}}

\newcommand{\Ham}{H^{\varepsilon}_{\rm BF}}

\newcommand{\be}{\begin{equation}}
\newcommand{\ee}{\end{equation}}
\newcommand{\er}{\hfill$\diamondsuit$}

\newtheorem{theorem}{Theorem}
\newtheorem{lemma}[theorem]{Lemma}
\newtheorem{proposition}[theorem]{Proposition}
\newtheorem{corollary}[theorem]{Corollary}

\theoremstyle{definition}
\newtheorem{definition}[theorem]{Definition}
\newtheorem{remark}[theorem]{Remark}

\begin{document}

\title{\LARGE\bf Effective dynamics
 for Bloch electrons:
 Peierls substitution and beyond}

\author{ \large Gianluca Panati, Herbert Spohn, Stefan Teufel \medskip  \\
\normalsize Zentrum Mathematik and Physik Department,\\
\normalsize Technische Universit\"{a}t M\"{u}nchen, 85747 Garching, Germany   \medskip\\
\normalsize email: panati@ma.tum.de, spohn@ma.tum.de,
teufel@ma.tum.de }

\date{May 16, 2003}
 
\maketitle

\begin{abstract}
We consider an electron moving in a periodic potential and subject
to an additional slowly varying external electrostatic potential,
$\phi(\epsi x)$, and vector potential $A(\epsi x)$, with $x \in
\R^d$ and $\epsi \ll 1$. We prove that associated to an isolated
family of Bloch bands there exists an almost invariant subspace
of $L^2(\R^d)$ and an effective Hamiltonian governing the
evolution inside this subspace to all orders in $\epsi$. To leading
order the effective Hamiltonian is given through the Peierls
substitution. We explicitly compute the first order correction.
From a semiclassical analysis of this effective quantum
Hamiltonian we establish the first order correction to the
standard semiclassical model of solid state physics.
\end{abstract}

\tableofcontents

\section{Introduction}

A central problem of solid state physics is to understand the
motion of electrons in the periodic potential generated by the
ionic cores. While the dynamics is quantum mechanical, many
electronic properties of solids can be understood already in the
  semiclassical approximation
\cite{AsMe,Ko,Za}. One argues that for suitable wave packets,
which are spread over many lattice spacings, the main effect of a
periodic potential   $V_{\Gamma}$ on the electron dynamics
consists in changing the dispersion relation from the free
kinetic energy $E_{\rm free}(k) = \frac{1}{2}\, k^2$ to the
modified kinetic energy $E_n(k)$ given by the $n^{\rm th}$ Bloch
band. Otherwise the electron responds to slowly varying external
potentials $A$, $\phi$ as in the case of a vanishing periodic
potential. Therefore the semiclassical equations of motion read
\be
\label{semiclassical_dynamics} \dot r = \nabla  E_n(\kappa) \,,\qquad
\dot \kappa = -\nabla \phi(r) + \dot r \vp B(r)\,, 
\ee 
where $r \in \R^3$ is the position of the electron, $\kappa = k - A(r)$ its kinetic momentum with 
$k$ its Bloch momentum, $-\nabla \phi$ the external electric field, and $B= \nabla \times A$
the external magnetic field. Note that there is a semiclassical
evolution for each Bloch band separately. (We choose units in
which the Planck constant $\hbar$, the speed $c$ of light, and the
mass $m$   of the electron are equal to one, and absorb the charge
$e$ into the potentials).

One goal of this article is to understand on a mathematical level
how these semiclassical equations emerge from the underlying
Schr\"odinger equation
\begin{eqnarray}
 \label{basic_dynamics}
 \I\,\epsi\, \partial_t\,\psi(x,t)  &=& \left( {\textstyle\frac{1}{2}}\big(-\I
\nabla_x - A(\epsi x)\big)^2 + V_{\Gamma}(x) + \phi(\epsi
x)\right) \psi(x,t) \nonumber \\
 &=& H^{\epsi}\psi(x,t)
\end{eqnarray}
in the limit $\epsi\to 0$ at leading order. Here the potential
$V_{\Gamma}:\R^3\to \R$ is periodic with respect to some regular
lattice $\Gamma$. $\Gamma$ is generated through the basis
$\{\gamma_1,\gamma_2,\gamma_3 \}$, $\gamma_j\in\R^3$, i.e.\
\[
\Gamma =\Big\{ x\in\R^3: x=
\textstyle{\sum_{j=1}^3}\alpha_j\,\gamma_j \,\,\,\mbox{for
some}\,\,\alpha \in \mathbb{Z}^3 \Big\}\,,
\]
and $V_{\Gamma}(x + \gamma) = V_{\Gamma}(x)$ for all
$\gamma\in\Gamma, x \in \R^3$. The spacing of the lattice $\Gamma$
defines the microscopic spatial scale. The external potentials
$A(\epsi x)$ and $\phi(\epsi x)$, with $A:\R^3\to\R^3$ and
$\phi:\R^3\to\R$, are slowly varying on the scale of the lattice,
as expressed through the  dimensionless scale parameter $\epsi$,
$\epsi\ll 1$. In particular, this means that the external fields
are weak compared to the fields generated by the ionic cores, a
condition which is
 satisfied for real metals  even for the strongest external electrostatic fields
available and for a wide range of  magnetic fields, see
\cite{AsMe}, Chapter~12.

In solid state physics the derivation of the semiclassical model
(\ref{semiclassical_dynamics}) received a lot of attention during
the 1950s to the 1970s. We mention representatively   the work
by Luttinger \cite{Lu}, Kohn \cite{Ko}, Blount \cite{Bl1,Bl2}
and Zak \cite{Za}. As late as 1962 Wannier \cite{Wa} argues
that the derivation of (\ref{semiclassical_dynamics}) from
(\ref{basic_dynamics}) is still incomplete.

On the mathematical side the semiclassical asymptotics  of the spectrum
 of $H^{\epsi}$ have been studied in great
detail by G\'{e}rard, Martinez and Sj\"{o}strand \cite{GMS} with predecessors 
\cite{BeRa,Bu,HeSj,Ne}. The large time asymptotics of the solutions to
(\ref{basic_dynamics}) without external potentials is studied in \cite{AsKn}
and the scattering theory is developed in \cite{GeNi}. However for the dynamics of wave functions,
our interest here, the results are modest. In \cite{GMMP} the case
$\phi=0, A=0$ is considered, in \cite{HST} and  \cite{BMP} a
proof is given for $A=0$, which leaves out many interesting
applications. The method of Gaussian beams is developed in \cite{GRT}
for a weak uniform magnetic field and in \cite{DGR} for magnetic Bloch bands.

In fact, as our title indicates, we are more ambitious and plan to
derive also the first order correction to
(\ref{semiclassical_dynamics}). The electron acquires then  a
$k$-dependent electric moment $\mathcal{A}_n(k)$ and magnetic
moment $\mathcal{M}_n(k)$. If the $n^{\rm th}$ band is
nondegenerate (hence isolated) with Bloch eigenfunctions
$\psi_n(k,x)$, the electric dipole moment is given by the Berry
connection 
\be 
\A_{n} (k) = \I \,\big\langle \psi_{n}(k), \nabla_k
\psi_{n}(k) \big\rangle 
\ee 
and the magnetic moment by the
Rammal-Wilkinson term
 \be 
 \mathcal{M}(k)_{n} =
\textstyle{\frac{\I}{2}}\,\big\langle \nabla_k\psi_{n}(k), \ \vp
(H_{\rm per}(k) - E(k))\nabla_k\psi_{n}(k) \big\rangle\,. 
\ee
Here $\langle \cdot \, , \cdot \rangle \, $ denotes the inner
product in $L^2(\R^3/\Gamma)$ and $H_{\rm per}(k)$ is $H^\epsi$ of
(\ref{basic_dynamics}) with $\phi = 0 = A$ for fixed Bloch
momentum $k$, see Eq.\ (\ref{H(k)}). As  will be explained in
detail, the corrected semiclassical equations read
\begin{eqnarray}
\dot r &=& \nabla_{\kappa} \Big( E_n(\kappa) - \epsi \, B(r)\cdot \M_n(\kappa)\Big)
- \epsi\, \dot  \kappa \vp \Omega_n(\kappa)\,,\nonumber\\\label{Semi1}\\
\dot  \kappa  &=&  -\nabla_r \Big(\phi(r) - \epsi\, B(r)\cdot
\M_n(\kappa)\Big) +\dot r \vp B(r) \nonumber
\end{eqnarray}
with $\Omega_n(k) = \nabla \vp \A_n(k)$ the curvature of the Berry
connection.

The issue of first order corrections to the semiclassical
equations of motion has been investigated recently by Sundaram and
Niu \cite{SuNi} in the context of magnetic Bloch bands, see also
Chang and Niu \cite{ChNi}. One adds in (\ref{basic_dynamics}) a
strong uniform magnetic field $B_0$, i.e. the vector potential
$A_0(x) = \frac{1}{2} B_0 \vp x$. If its magnetic flux per unit
cell is rational, then the Hamiltonian in (\ref{basic_dynamics})
is still periodic at the expense of a larger unit cell and
replacing the usual translations by the magnetic
translations. Equation (\ref{Semi1}) remains formally unaltered,
only $E_n$ now refers to the energy of the magnetic subband.
Instructive plots of $\Omega_n$ and $\M_n$ are provided in \cite{SuNi}
for the particular case of the $2$-dimensional Hofstadter  model
at rational flux $1/3$.   The first order corrections obtained
in \cite{SuNi} agree with our equation (\ref{Semi1}), except for
the term of order $\epsi$ in the second equation. On a technical
level   magnetic Bloch bands require some extra
considerations and we defer them to a forthcoming paper \cite{PST3}.

It has been recognized repeatedly, as e.g.\ emphasized in
\cite{ABL}, that the geometric phases appearing in the first order correction
contain novel physics as compared to the leading order. Bloch
electrons are no exception. For example for the case of magnetic Bloch bands,
the equations of motion (\ref{Semi1}) 
provide a simple semiclassical explanation of the quantum Hall
effect. Let us specialize (\ref{Semi1}) to two dimensions and take $B(r)=0$,
$\phi(r) = - \mathcal{E} \cdot r$, i.e.\ a weak driving electric
field and a strong uniform magnetic field with rational flux. Then, since $\kappa=k$,
the equations of motion   become $\dot r = \nabla_k
E_n(k) + \mathcal{E}^\perp\Omega_n(k)$, $\dot k = \mathcal{E}$,
where $\Omega_n$ is now scalar, and $\mathcal{E}^\perp$ is
$\mathcal{E}$ rotated by $\pi/2$. We assume initially $k(0)=k$ and
a completely filled band, which means to integrate with respect to
$k$ over the first Brillouin zone $M^*$. Then the average current for
band $n$ is given by
\[
j_n = \int_{M^{*}} \D k \,\dot r(k) = \int_{M^{*}} \D k
\,\big(\nabla_{ k } E_n( k )-     \mathcal{E}^\perp \Omega_n( k
)\big) =  - \mathcal{E}^\perp \int_{M^{*}} \D k\,\Omega_n( k)\,.
\]
$\int_{M^{*}} \D k\,\Omega_n( k)$ is the Chern number of the
magnetic Bloch bundle and as such an integer. Further applications
related to the semiclassical first order corrections are the anomalous
Hall effect \cite{JNM} and the thermodynamics of the Hofstadter
model \cite{GaAv}.

Our derivation of (\ref{Semi1}) from (\ref{basic_dynamics})
proceeds in two conceptually and mathematically distinct steps.
The first step is to obtain an effective Hamiltonian whose unitary
group closely approximates the solution to the Schr\"{o}dinger
equation (\ref{basic_dynamics}) for $\epsi$ small, in case
the initial wave function lies in a subspace corresponding to a prescribed 
family of Bloch bands. Inside the family, band crossings
and almost crossings are allowed. It is crucial however that for
every $k$ the family of bands is separated by a gap from the
remaining energy bands. Then, associated to the given family of
bands, there is a subspace $\Pi^\epsi L^2(\R^3)$ which is adiabatically
decoupled from its orthogonal complement to all orders in $\epsi$.
The effective Hamiltonian generates the approximate time evolution
in $\Pi^\epsi L^2(\R^3)$. 

Compared to the space-adiabatic perturbation theory
developed in \cite{PST1}, as a new element we have to face the
fact that the classical phase space is $(\R^3/ \Gamma^{*}) \times
\R^3 $, $\Gamma^{*}$ the lattice dual to $\Gamma$ and $\R^3/
\Gamma^{*} =M^*$ the first Brillouin zone. To come close
to the scheme in \cite{PST1} a natural approach is to use the
extended zone scheme. Going from one cell to the next, one picks up
a phase factor which necessitates to generalize the
pseudodifferential calculus to $\tau$-equivariant symbols, see
Appendix~A.

The effective Hamiltonian is expanded in an $\epsi$-independent
reference Hilbert space. For example, for a nondegenerate band the reference space
is $L^2(M^*,\D k)$ and the leading order    effective Hamiltonian is given
through the Peierls substitution 
\be 
h_0(k,\I\epsi\nabla_k) = E_n(k
-A(\I \epsi \nabla_k)) + \phi(\I \epsi \nabla_k) \,,
\ee 
  where $\I \nabla_k$ is understood
with periodic boundary conditions on $M^*$. 

The natural second step consists in a semiclassical analysis of
the effective Hamiltonian. It is a standard result that the
unitary group generated by $h_0$ is well approximated by the
semiclassical equations (\ref{semiclassical_dynamics}). At next
order, $h_0(k,\I\epsi\nabla_k)$ is corrected to
$h_0(k,\I\epsi\nabla_k) + \epsi h_1(k,\I\epsi\nabla_k)$, with $h_1$
given in (\ref{h1 special}). However (\ref{Semi1}) is \emph{not}
the semiclassical evolution corresponding to that Hamiltonian. The
reason is that the  subspace $\Pi^\epsi L^2(\R^3)$ is mapped   to the reference Hilbert space
$L^2(M^*,\D k)$ through a unitary operator which itself depends on $\epsi$.
Therefore, the transformation of observables
generates an $\epsi$-dependence in addition to the transformation of time-evolved states.
 If done properly, one arrives at (\ref{Semi1}).

To give a brief outline of the paper. In Section~2 we discuss the
periodic Hamiltonian. In particular we recall the unitary Zak
transform and state our assumptions on  $V_{\Gamma}$, $A$, $\phi$ and
the gap condition. In Section~3 we apply the  space-adiabatic
perturbation theory to the present case, using the
pseudodifferential calculus developed in Appendix~A. 
The semiclassical analysis of the effective Hamiltonian
including first order is carried out in Section~4. The precise link between
(\ref{basic_dynamics}) and (\ref{Semi1}) is stated in
Theorem~\ref{EgCor}. In Appendix~B we show that the equations (\ref{Semi1}) are of Hamiltonian form with respect to an appropriate
symplectic structure.

\section{The periodic  Hamiltonian}

In order to formulate our setup we first
need to recall several well known facts about the periodic
Hamiltonian
\[
H_{\rm per} := -\frac{1}{2} \Lap + V_\Gamma\,,
\]
acting in $L^2(\R^d)$, keeping from now on the dimension $d$ arbitrary. The
potential $V_\Gamma$ is periodic with respect to the lattice
$\Gamma$. Its dual lattice $\Gamma^*$ is defined as the lattice
generated by the dual basis $\{\gamma_1^*,\ldots,\gamma_d^*\}$
determined through the conditions $\gamma_i\cdot\gamma_j^* = 2\pi
\delta_{ij}$, $i,j\indexd$. The centered \ix{fundamental domain}
fundamental domain of $\Gamma$ is denoted by
\[
 M = \Big\{ x\in\R^d: x=
\textstyle{\sum_{j=1}^d}\alpha_j\,\gamma_j \,\,\,\mbox{for}\,\,\alpha_j\in
[-\textstyle{\frac{1}{2},\frac{1}{2}}]
 \Big\}\,,
\]
 and analogously the centered fundamental domain of $\Gamma^*$ is
denoted by $M^*$. In solid state physics the set $M^*$ is called
\ix{Brillouin zone} the {\em first Brillouin zone}. In the
following $M^*$ is always equipped with the {\em normalized}
Lebesgue measure denoted by $\D k$. We introduce the notation
$x=[x] +\gamma$ for the a.e.\ unique decomposition of $x\in\R^d$
as a sum of $[x]\in M$ and $\gamma\in\Gamma$. We use the same
brackets for the analogous splitting $k=[k] + \gamma^*$.

We employ a variant of the Bloch-Floquet transform,
called the Zak transform (also Lifshitz-Gelfand-Zak
transform). The Zak transform of a function $\psi\in\Sch(\R^d)$ is
defined as \be \label{BFdef} (\U\psi)(k,x):=\sum_{\gamma\in\Gamma}
\E^{-\I (x+\gamma)\cdot k}\psi(x+\gamma),\,\,\, (k,x)\in\R^{2d},
\ee and one directly reads off from (\ref{BFdef}) the following
periodicity properties
 \be
\label{BF1} \big(\U\psi\big) (k, y+\gamma) = \big( \U\psi\big)
(k,y)\quad \mbox{ for all} \quad \gamma\in\Gamma\,,
\ee
\be
\big(\U\psi\big) (k+\gamma^*, y) = \E^{-\I
y\cdot\gamma^*}\,\big( \U\psi\big) (k,y) \quad\mbox{ for all}
\quad \gamma^*\in\Gamma^*\,. \label{BF2}
\ee

\noindent From (\ref{BF1}) it follows that, for any fixed
$k\in{\R^d}$, $\big( \U\psi \big)(k,\cdot)$ is a $\Gamma$-periodic
function and can thus be regarded as an element of $L^2(\T^d)$,
$\T^d$ being the flat torus $\R^d/\Gamma$. Equation (\ref{BF2})
involves a unitary representation of the group of lattice
translations on $\Gamma^*$ (denoted again as $\Gamma^*$ with a
little abuse of notation), given by
\[
\tau:\Gamma^*\to\U(L^2(\T^d))\,,\quad\gamma^*\mapsto
\tau(\gamma^*)\,, \quad (\tau(\gamma^*)\ph)(y) =
\E^{\I\,y\cdot\gamma^*} \ph(y).
\] It will turn out convenient to introduce the Hilbert space \be
\Hi_\tau :=\Big\{ \psi\in L^2_{\rm loc}(\R^d, L^2(\T^d)):\,\,
\psi(k -\gamma^*) = \tau(\gamma^*)\,\psi(k) \Big\}\,, \label {H
tau} \ee equipped with the inner product
\[
\langle \psi,\,\ph\rangle_{\Hi_\tau} = \int_{M^{*}}\D k\, \langle
\psi(k),\,\ph(k)\rangle_{L^2(\mathbb{T})}\,.
\]
Notice that if one considers the trivial representation, i.e.\
$\tau \equiv {\bf 1}$, then $\Hi_\tau$  is simply a space of
$\Gamma^{*}$-periodic vector-valued functions over $\R^d$.

Obviously, there is a natural isomorphism  between $\Hi_\tau$ and
$L^2(M^{*},L^2(\T^d))$ given by restriction from $\R^d$ to
$M^{*}$, and with inverse given by $\tau$-equivariant
continuation, as suggested by (\ref{BF2}). The reason for working
with $\Hi_\tau$ instead of $L^2(M^{*},L^2(\T^d))$ is twofold.
First of all it allows to apply the pseudodifferential calculus as
developed in Appendix A. On the other hand it makes statements
about domains of operators more transparent as we shall see.

The map defined by (\ref{BFdef}) extends to a unitary operator
\[
\U: L^2(\R^d)\to \Ht \cong L^2(M^{*}, L^2(\T^d)) \cong
L^2(M^{*})\otimes L^2(\T^d)\,.
\]
  $\U$ is an isometry and   $\U^{-1}$ given
through 
\be\label{BFinv} \big(\U^{-1}\ph\big)(x) = \int_{M^{*}}\D
k\, \,\E^{\I x\cdot k}\,\ph(k,[x]) 
\ee 
satisfies $\U^{-1}\U\psi =
\psi$ for $\psi\in\Sch(\R^d)$, as can be checked by direct
calculation.   $\U^{-1}$
extends to an isometry from $\Ht$ to $L^2(\R^d)$. Hence $\U^{-1}$
must be injective and as a consequence  $\U$ must be surjective,
thus unitary.

In order to determine the Zak transform of operators like the
full Hamiltonian in (\ref{basic_dynamics}), we need to discuss how
differential and multiplication operators behave under  the Zak
transform, see \cite{Bl1}, \cite{Za}. Let $P=-\I \nabla_x$ with
domain $H^1(\R^d)$ and $Q$ multiplication by $x$ on the maximal
domain. Then
\begin{eqnarray}\label{P trasf}
\U \,P\, \U^{-1} &=& {{\bf 1}}\otimes -\I \nabla_y^{\rm per} + k\otimes{{\bf 1}} \,,\\
\U \,Q\, \U^{-1} &= & \I\nabla^\tau_k   \,,\label{Q trasf}
\end{eqnarray}
where  $-\I  \nabla_y^{\rm per}$ is equipped with periodic
boundary conditions or, equivalently, operating on the domain
$H^1(\T^d)$. The domain of  $\I\nabla^\tau_k$ is $\Hi_\tau\cap
H^1_{\rm loc}(\R^d, L^2(\T^d))$, i.e.\ it consists of
distributions in  $H^1(M^{*}, L^2(\T^d))$ which satisfy the
$y$-dependent boundary condition associated with (\ref{BF2}). In
addition to (\ref{P trasf}) and (\ref{Q trasf}) we notice that
multiplication with a $\Gamma$-periodic function like $V_\Gamma$
is mapped into multiplication with the same function, i.e.\
 $\U\,V_{\Gamma}(x)\,\U^{-1} = {\bf 1} \otimes V_\Gamma(y)$.

For later use we remark that the following relations can be
checked using the definitions (\ref{BFdef}) and (\ref{BFinv}),
\begin{eqnarray*}
\psi\in H^m(\R^d)\,,\,\,m \geq 0 &\quad\Longleftrightarrow\quad &
\U\psi \in L^2(B,H^m(\T^d))\,, \\
\langle x\rangle^m\psi(x)\in L^2(\R^d) \,,\,\,m\geq 0
&\quad\Longleftrightarrow\quad & \U\psi \in \Hi_\tau\cap H^m_{\rm
loc}(\R^d,L^2(\T^d)) \,.
\end{eqnarray*}

\begin{remark}  The Bloch-Floquet transform is usually defined
as \be \label{BFdef2}
(\widetilde\U\psi)(k,x):=\sum_{\gamma\in\Gamma} \E^{-\I  x \cdot
k}\psi(x+\gamma),\,\,\, (k,x)\in\R^{2d}\,. \ee for
$\psi\in\Sch(\R^d)$. In contrast to (\ref{BFdef}),  functions in
the range of\ $\widetilde \U$ are periodic in $k$ and
quasi-periodic in $y$,
 \be
\label{BF12} \big(\widetilde\U\psi\big) (k, y+\gamma) = \E^{\I
k\cdot\gamma}\,\big( \widetilde\U\psi\big) (k,y)\quad \mbox{ for
all} \quad \gamma\in\Gamma\,, \ee \be \big(\widetilde\U\psi\big)
(k+\gamma^*, y) = \big( \widetilde\U\psi\big) (k,y) \quad\mbox{
for all} \quad \gamma^*\in\Gamma^*\,. \label{BF22} \ee Our choice
of using the Zak transform $\U$ instead of $\widetilde \U$ comes
from the fact that the transform of the gradient has a domain
which is independent of $k\in M^{*}$, see (\ref{P trasf}). As we shall see, this is 
essential for the application of the
pseudodifferential calculus of   Appendix~A.\er
\end{remark}

 For the  Zak transform of the free Hamiltonian one finds
 \[
 \U\,H_{\rm per}\,\U^{-1} = \int_{M^{*}}^\oplus\D k\,H_{\rm
per}(k)
\]
with 
\be \label{H(k)} H_{\rm per}(k) = \frac{1}{2}\big( -\I
\nabla_y + k\big)^2 + V_\Gamma(y)\,,\quad k\in M^{*} \,.
\ee 
For fixed
$k\in M^{*}$ the operator $H_{\rm per}(k)$ acts on $L^2(\T^d)$
with domain $H^2(\T^d)$ independent of $k\in M^{*}$, whenever the
following assumption on the potential is satisfied.\\

 \noindent {\bf Assumption A$_1$.}\quad {\em We assume that $V_\Gamma$ is
infinitesimally bounded with respect to $-\Lap$ and that $\phi \in
C^\infty_{\rm b}(\R^d,\R)$ and $A_j\in C^\infty_{\rm b}(\R^d,\R)$
 for
any  $j\in\{ 1,\ldots,d \} $.}\\
 
Here $C^\infty_{\rm b}(\R^d,\R)$ denotes the space of bounded
smooth functions with derivatives of any order bounded. From this
assumption it follows in particular that also the full Hamiltonian
$H^\epsi$ of (\ref{basic_dynamics}) is self-adjoint on
$H^2(\R^d)$. Assumption (A$_1$) excludes the case of globally
constant electric and magnetic field. However, since we are not concerned 
with the spectral analysis of $H^\epsi$, but with the dynamics of states for 
large but finite times, locally constant fields serve us as well.

The band structure of the fibered spectrum of $H_{\rm per}$ is
crucial for the following. The resolvent $R_\lambda^0 = (H_0(k)
-\lambda)^{-1}$ of the
 operator $H_0(k)=\frac{1}{2}\big( -\I \nabla_y + k\big)^2$
is compact  for fixed $k\in M^{*}$. Since, by assumption,
$R_\lambda V_\Gamma$ is bounded, also $R_\lambda = (H_{\rm per}(k)
-\lambda)^{-1} = R_\lambda^0 + R_\lambda V_\Gamma R_\lambda^0$ is
compact. As a consequence $H_{\rm per}(k)$ has purely discrete
spectrum with eigenvalues
 of finite multiplicity which  accumulate at infinity.
  A more detailed discussion can be
found e.g.\ in \cite{Wi}.
For definiteness the eigenvalues are enumerated increasingly as
$E_1(k) \leq E_2(k) \leq E_3(k)\leq\ldots$ and repeated according
to their multiplicity. The corresponding normalized eigenfunctions
$\{\ph_n(k)\}_{n\in\N}\subset H^2(\T^d)$ are called Bloch
functions \ix{Bloch functions} and form, for any fixed $k$, an
orthonormal basis of $L^2(\T^d)$. We will call $E_n(k)$ the
$n^{\rm th}$ band function or just the $n^{\rm th}$ band. \ix{Bloch bands} Notice that, with
this choice of the labelling, $E_n(k)$ and $\varphi_n(k)$ are
generally \emph{not} smooth functions of $k$ due to eigenvalue
crossings. Since \be H_{\rm per}(k-\gamma^*) =
\tau(\gamma^*)\,H_{\rm per}(k) \,\tau(\gamma^*)^{-1}\,,
\label{Hper equiv} \ee the band functions $E_n(k)$ are periodic
with respect to $\Gamma^*$.

\begin{definition}\label{Isodef}
A family of Bloch bands $\{ E_n(k)\}_{n\in\mathcal I}$, $\mathcal
I = [I_{-}, I_{+}] \bigcap \N $, is called   \textbf{isolated}, or
satisfies the \textbf{gap condition}, if
\[
\inf_{k\in M^{*}} {\rm dist} \Big(\, \bigcup_{n\in\mathcal{I}}
\{E_n(k) \},\, \bigcup_{m\notin\mathcal{I}} \{E_m(k)\}\,\Big)=:
C_{\rm g}>0\,.
\]
\end{definition}

In the following we fix an index set $\mathcal{I}\subset\N$
for an isolated family of bands. Let $P_{\mathcal{I}}(k)$
be the spectral projector of $H_{\rm per}(k)$ corresponding to the
eigenvalues $\{ E_n(k)\}_{n\in\mathcal I}$, then $P_{\mathcal{I}}
:= \int^\oplus_{M^{*}}\D k\, P_{\mathcal{I}}(k)$ defines  the projector
on the given isolated family of bands.
In terms of Bloch functions  
 $P_{\mathcal{I}}(k)= \sum_{n \in \mathcal{I}}
|\ph_n(k)\rangle\langle\ph_n(k)|$. However, in general, $\ph_n(k)$
are not smooth functions of $k$ at eigenvalue crossings, while $P_{\mathcal{I}}(k)$ is
 a smooth
function of $k$ because of the gap condition. Moreover, from (\ref{Hper equiv}) it follows that
\[
 P_{\mathcal{I}}(k-\gamma^*) = \tau(\gamma^*)\,
P_{\mathcal{I}}(k) \,\tau(\gamma^*)^{-1}\,.
\]

For the mapping to the reference space we will need the following
assumption.\\

\noindent{\bf Assumption A$_2$.}\quad {\em \hspace{-1.3pt}If the
 isolated family of bands
$\{E_n\hspace{-1pt}(k)\}_{n\in\mathcal I}$ is degenerate, in the
sense that $\ell=|\mathcal{I}| >1$, then  we assume that there
exists an orthonormal basis  $\left\{ \psi _{j}(k)
\right\}_{j=1}^{\ell}$ of ${\rm Ran}P_{\mathcal{I}}(k)$ whose
elements are smooth and
 $\tau$-equivariant with respect to $k$, i.e.\
 $\psi_j(k-\gamma^*)= \tau(\gamma^*) \psi_j(k)$ for all
 $j\in\{1,\ldots,\ell\}$ and $\gamma^*\in\Gamma^*$.}\\

In the case   of a single isolated
$\ell$-fold degenerate Bloch band (i.e.\ $E_n(k)= E_{*}(k)$ for
every $n \in \mathcal{I}, \, |\mathcal{I}|= \ell $), Assumption
(A$_2$) is equivalent to the existence of an orthonormal basis
consisting of smooth and $\tau$-equivariant Bloch functions. On
the other side, if there are \textit{eigenvalue crossings} inside
the family of bands, Assumption (A$_2$) requires only that
$\psi_{j}(k)$ is an eigenfunction of the corresponding
eigenprojection $P_{\mathcal{I}}(k)$ and not of  the free
Hamiltonian $H_{\rm per}(k)$.

From the geometrical viewpoint Assumption (A$_2$) is equivalent to
the triviality of a complex vector bundle over $\T^d$, namely the
bundle of the  null spaces of $ 1 - P_{\mathcal{I}}(k)$ for $k \in
M^{*}$. In this geometrical perspective it is not difficult to see
that Assumption (A$_2$) is always satisfied if either $d=1$ or
$\ell =1$. Indeed, classification theory for bundles implies that
any complex vector bundle over $\T^1 = S^1$ is trivial. As for
$\ell =1$, it is a classical result, due to Kostant and Weil, that
smooth complex \emph{line} bundles are completely classified by
their first integer Chern class. In our case, the time-reversal
symmetry of $H_{\rm per}$ implies the vanishing of the first
integer Chern class, and therefore the triviality of the bundle.
The same, and indeed slightly stronger, results can be proved with
analytical techniques, as in \cite{Ne} and references therein. By
pushing forward the geometrical approach  above, we expect that
Assumption (A$_2$) is generically satisfied for $d \leq 3$, as it
will be discussed in \cite{Pa}.

In the presence of a strong external magnetic field the Bloch
bands split into magnetic sub-bands. Generically, their first
Chern number does not vanish and therefore Assumption (A$_2$)
fails. As well understood and discussed in the introduction,
 the nonvanishing of the first Chern
number is directly linked to the integer quantum Hall effect
\cite{TKNN,Si}, hence our interest in extending Theorem 3 to
magnetic Bloch bands. The required modifications of our theory
will be discussed in  \cite{PST3}.

 \section{Space-adiabatic perturbation for Bloch bands}

Let $P_n(k)=  |\ph_{n}(k)\rangle\langle\ph_{n}(k)|$. Then the
projector on the $n^{\rm th}$ band subspace is given through $P_n
= \int^\oplus_{M^*}\D k\, P_n(k)$. By construction the band
subspaces are invariant under the dynamics generated by $H_{\rm
per}$,
\[
\Big[ \,\E^{-\I  \U H_{\rm per}\U^{-1}\,s},\,P_n\,\Big] = \Big[
\,\E^{-\I  E_n(k) s},\,P_n\,\Big] = 0\quad\mbox{for all}
\,\,n\in\N\,,\,\, s\in\R\,.
\]
Notice that $P_n$ is not a spectral projector of $H_{\rm per}$, in
general, since in more than one space dimension it can happen that
e.g.\ $E_n(k) < E_{n+1}(k)$ for all $k\in M^*$ but $\inf_k E_{n+1}(k) < \sup_k E_n(k)$. 
According to the identity  (\ref{P trasf}),  in the
original representation $H_{\rm per}$ acts on the $n^{\rm th}$
band subspace as
\[
H_{\rm per} \psi = \U^{-1} (E_n(k)\otimes{\bf 1})\U \,\psi =
E_n(-\I \nabla_x)\,\psi\,,
\]
where $\psi\in \U^{-1}P_n\,\U \,L^2(\R^d)$. In other words, under
the time evolution generated by the periodic Hamiltonian wave
functions in the $n^{\rm th}$ band subspace propagate freely but
with a modified dispersion relation given through the  $n^{\rm
th}$ band function $E_n(k)$.

In the presence of non-periodic external fields the subspaces
$P_n\Ht$ are no longer invariant, since the external fields induce
transitions between different band subspaces. If the potentials
are varying slowly, these transitions are small and one expects
that there  still exist almost invariant subspaces associated with
isolated Bloch bands. To construct them, and to study the dynamics
inside these almost invariant subspaces, we apply adiabatic
perturbation  to perturbed Bloch bands.

We first present a theorem which summarizes the main results of
this section. The remaining parts   give the results and the
proofs of the three main steps in space-adiabatic perturbation
theory: In Section 3.1 we construct the almost invariant subspaces
associated with isolated Bloch bands. In Section 3.2 we explain
how to unitarily map the decoupled subspace  to a suitable
reference Hilbert space. In this reference space the action of the
full Hamiltonian is given through a semiclassical
pseudodifferential operator, whose power series expansion can be
computed to any order in $\epsi$. This effective Hamiltonian is
constructed in Section 3.3 and we compute explicitly  its
principal and subprincipal symbol. The main technical innovation
necessary in order to apply the scheme to the present case is the
development of a pseudo\-differential calculus for operators
acting on sections of a bundle over the flat torus $M^{*}$, or,
equivalently, acting on the space $\Ht$. This task is deferred to
Appendix A.

Before going into the details of the construction we present a
theorem which encompasses the main results of this section.
Generalizing from (\ref{H tau}) it is convenient to introduce the
following notation. For any separable Hilbert space $\Hi_{\rm f}$
and any unitary representation $\tau:\Gamma^*\to\U(\Hi_{\rm f})$,
one defines the Hilbert space
\[
L^2_\tau(\R^d, \Hi_{\rm f})
:=\Big\{ \psi\in L^2_{\rm loc}(\R^d, \Hi_{\rm f}):\,\, \psi(k
-\gamma^*) = \tau(\gamma^*)\,\psi(k) \Big\}\,,
\]
equipped with
the inner product
\[
\langle \psi,\,\ph\rangle_{L^2_\tau} = \int_{M^{*}}\D k\, \langle
\psi(k),\,\ph(k)\rangle_{\Hi_{\rm f}}\,.
\]
Using the results of the previous section and imposing Assumption
(A$_1$), the Zak transform of the full Hamiltonian in
(\ref{basic_dynamics}) is given through \be H^\epsi_{\rm Z} :=
\U\,H^\epsi\,\U^{-1} = \frac{1}{2}\Big( -\I \nabla_y + k -
A\big(\I\epsi\nabla_k^\tau\big)\Big)^2 + V_\Gamma(y) +
\phi\big(\I\epsi\nabla_k^\tau\big) \label{Ham BF} \ee with domain
$L^2_{\tau}(\R^d, H^2(\T^d))$. 

The application of space-adiabatic perturbation theory  to an
isolated family of bands $\{E_n(k)\}_{n\in\mathcal I}$ yields the following
result, where the reference Hilbert space for the effective
dynamics is
 $\K := L^2(M^{*}) \otimes \C^\ell $ with $\ell:= {\rm dim} P_{\mathcal{I}}(k)$.

\begin{theorem}[Peierls substitution and higher order corrections]
\label{Th main}

\noindent Let $\{E_n\}_{n\in\mathcal{I}}$ be an isolated family of
bands, see\ Definition \ref{Isodef}, and let the Assumptions
(A$_1$) and (A$_2$) be satisfied. Then there exist
\begin{enumerate}
\item
 an orthogonal
projection $\Pi^\epsi \in \B(\Ht)$,
\item
a unitary map $U^\epsi \in \B(\Pi^\epsi\Ht,\K)$, and
\item
a self-adjoint operator $ \widehat{h}  \in \B(\K)$
\end{enumerate}
 such that
\[
\big\|\,\big[ \, H^\epsi_{\rm Z}, \,\Pi^\epsi \,\big] \,\big\|= \Or(\epsi^\infty)
\,, \qquad \|\,\Pi^\epsi - P_\mathcal{I} \,\|=\Or(\epsi)
 \]
and
\[
\big\|\,\big( e^{-i H^\epsi_{\rm Z} t} -U^{\epsi\,*}\ e^{-i \widehat{h}  t}\ U^\epsi
\big) \Pi^\epsi\, \big\|=\Or (\epsi ^{\infty }(1+|t|))\,.
\]
The effective Hamiltonian  $\widehat{h} $ is the Weyl quantization
of a semiclassical symbol $h  \in S^{1}_{\tau \equiv {\bf
1}}(\epsi, \B(\C^\ell))$ with    an asymptotic expansion   to any order. The $\B(\C^\ell)$-valued
principal symbol $h_0(k,r)$ has matrix-elements \be
h_0(k,r)_{\alpha \beta}= \big\langle \psi_\alpha(k-A(r)),
H_0(k,r)\, \psi_\beta(k-A(r))\big\rangle\,, \ee where $\alpha,
\beta \in \{1, \ldots, \ell \}$ and $H_0(k,r)$ is defined in
(\ref{H symbol}).
\end{theorem}

The general formula for the subprincipal symbol of the effective
Hamiltonian can be found in \cite{PST1}. The structure and the
interpretation of the effective Hamiltonian are most transparent
for the case of a single isolated band.

\begin{corollary} \label{MainCor}
For an isolated $\ell$-fold degenerate eigenvalue $E
(k)$  the $\B(\C^\ell)$-valued symbol $h (k,r) = h_0(k,r) + \epsi
h_1(k,r) + \Or_0(\epsi^2)$ constructed in Theorem~\ref{Th main} has matrix-elements
\be
 h_0(k,r)_{\alpha \beta}= \big( E(k - A(r)) + \phi(r)\big)
\delta_{\alpha \beta} \label{h0 specialbis}
\ee
 and
\begin{eqnarray}
\label{h1 special}
 \lefteqn{ \hspace{-.5cm}h_1(k,r)_{\alpha
\beta} = - \big( - \nabla\phi(r) + \nabla E(\widetilde k) \times
B(r)\big) \cdot \A (\widetilde k)_{\alpha \beta}- B(r)\cdot
\M(\widetilde k)_{\alpha
\beta}}\\ \nonumber\\
&:=& \Big( \partial_j \phi(r) -   \partial_l
E(\widetilde{k})\,\big( \partial_j A_l(r) -  \partial_l
A_j(r)\big)\Big) \, \A_j(\widetilde{k})_{\alpha\beta}
 \nonumber  \\
&&-   \big(\partial_j A_l  - \partial_l
A_j \big)(r)  \,\, {\rm Re} \left[ {\textstyle \frac{\I}{2}} \,  \big\langle \partial_l\psi_{\alpha}(\widetilde k),
  (H_{\rm per} - E)(\widetilde k) \
\partial_j \psi_{\beta} (\widetilde k) \big\rangle_{\Hi_{\rm f}}\right]\,,   \nonumber
\end{eqnarray}
 where  summation over indices appearing twice is implicit, $\widetilde k(k,r) = k - A(r)$, and $\alpha, \beta \in \{1, \ldots, \ell \}$.
  The  coefficients of the Berry connection are
\be\label{5AgeoDef}
 \A (k)_{\alpha \beta} = \I \,\big\langle \psi_{\alpha}(k),
\nabla \psi_{\beta}(k) \big\rangle_{\Hi_{\rm f}} \,.
\ee
 \end{corollary}

In dimension $d=3$ the subprincipal symbol (\ref{h1 special}) has
a straightforward physical interpretation. The 2-forms $B$ and
$\M$ are naturally identified with the vectors $B= {\rm curl} A$
and
\[
\M(k)_{\alpha \beta} =  \textstyle{\frac{\I}{2}}\,\big\langle
\nabla\psi_{\alpha}(k), \ \vp (H_{\rm per}(k) -
E(k))\nabla\psi_{\beta}(k) \big\rangle_{\Hi_{\rm f}}\,.
\]
Therefore the symbol of the effective Hamiltonian has the same
form as the energy of a classical charge distribution in weak
external fields, in first order multipole expansion. In this sense
 $\A (k)$ is interpreted as an effective electric dipole moment and $\M(k)$
 as an effective magnetic dipole moment.

\begin{remark}\label{NotRem}
Our results hold for arbitrary dimension $d$. However, to simplify presentation,
we use a notation motivated
by the vector product and the duality between 1-forms and 2-forms for $d=3$.
If $d\not= 3$, then $B$, $\Omega_n$ and $M_n$ are 2-forms.
The inner product of 2-forms is
\[
B \cdot M :=  *^{-1}(B \wedge *M)    =\sum_{j=1}^d \sum_{i=1}^d  B_{ij} M_{ij}\,,
\]
where $*$ denotes the Hodge duality induced by the euclidian metric,
and for a vector field $w$ and a 2-form $F$ the ``vector product'' is
\[
(w \vp F)_j := (*^{-1}(w \wedge *F))_j = \sum_{i=1}^d  w_i
F_{ij}\,,
\]
where the duality between 1-forms and vector fields was used implicitly.  \er
\end{remark}

Theorem \ref{Th main}  is a direct consequence of the results
proved in Propositions \ref{Prop Invariant subspace}, \ref{Prop
unitaries} and \ref{Prop Heff}. The proof of Corollary \ref{MainCor}
is given at the end of this section.

As mentioned before, the main idea of the proof is to adapt the
general scheme of space-adiabatic perturbation theory to the case
of the Bloch electron. While formally this seems straightforward,
one must overcome two mathematical problems. First of all,   in
the present case the symbols are \emph{unbounded}-operator-valued
functions. One can deal with unbounded-operator-valued symbols by
considering them as bounded operators from their domain equipped
with the graph norm into the Hilbert space, see e.g.\ \cite{DiSj}.
The second, more serious problem consists in setting up a Weyl
calculus for operators acting on  spaces like
$L^2_\tau(\R^d,\Hi_{\rm f})$. This is done in    Appendix~A and we
will use in this section the terminology and notations introduced
there.

The results of Appendix A allow us to write the Hamiltonian
$H^\epsi_{\rm Z}$ as the Weyl quantization
$H_0(k,\I\epsi\nabla_k)$ of the $\tau$-equivariant symbol 
\be
H_0(k,r)=\frac{1}{2}\big( -\I\nabla _{x}+k- A(r)\big)
^{2}+V_{\Gamma}(x)+ \phi(r) \label{H symbol} 
\ee 
acting on the
Hilbert space $\Hi_{\rm f}:=L^{2}(\T^{d},dx)$ with constant domain
$\Do := H^2(\T^d)$. For sake of   clarity, we spend two more words
on this point. For any fixed $(k,r) \in \R^{2d}$, $H_0(k,r)$ is
regarded as a bounded operator from $\Do$ to $\Hi_{\rm f}$ which
is $\tau$-equivariant with respect to the \emph{bounded}
representation $ \tau_{1} := \tau|_{\Do}$ acting on $\Do$ and the
{\em unitary} representation $ \tau_{2} := \tau$ acting on
$\Hi_{\rm f}$, see    Definition~\ref{ABDefequivsymbol}. Then the
general theory developed in Appendix A can be applied. The usual
Weyl quantization of $H_0$ is an operator from $\Sch'(\R^d, \Do)$
to $\Sch'(\R^d, \Hi_{\rm f})$ given by 
\be 
\widehat{H}_0 =
\frac{1}{2}\Big( -\I \nabla_y + k - A\big(\I\epsi\nabla_k
\big)\Big)^2 + V_\Gamma(y) + \phi \big(\I\epsi\nabla_k \big)
\label{Weyl H}\,. 
\ee
 Then $\widehat{H}_0$ can be restricted to $L^2_{\rm loc}(\R^d,
\Do)$, since $A$ and $\phi$ are smooth and bounded.  Since $H_0$ is
a $\tau$-equivariant symbol,   $\widehat{H}_0$ preserves
$\tau$-equivariance and can then be restricted to an operator from
$L^2_{\tau}(\R^d, \Do)$ to $L^2_{\tau}(\R^d, \Hi_{\rm f})$. To
conclude that (\ref{Weyl H}), restricted to $L^2_{\tau}(\R^d,
\Do)$, agrees with (\ref{Ham BF}), it is enough to recall that $\I
\nabla_k^{\tau}$ is defined as $\I \nabla_k$ restricted to $H^1
\cap \Ht$ and to use the spectral calculus.

Moreover, if one introduces the order function $w(k,r) :=
(1+k^2)$, then $H_0 \in S^w_\tau(\B(\Do,\Hi))$. More generally, we
will give the proofs for any symbol $H \in
S^w_\tau(\epsi,\B(\Do,\Hi))$, whose principal symbol is then
denoted by $H_0$.

\subsection{The almost invariant subspace}

In this section we construct the adiabatically decoupled subspace
associated with an isolated Bloch band. Similar constructions have
a considerable history and we refer to \cite{MaSo,NeSo,PST1,Te1} and
references therein.

Given an isolated family of bands $\{ E_n(k)\}_{n\in\mathcal I}$, we
define   $\pi_{0}(k,r)  = P_{\mathcal{I}}(k - A(r))$. It follows
from the $\tau$-equivariance of $H_0$ and from the gap condition
that $\pi_{0} \in S_{\tau}^{1}( \B(\Hi_{\rm f}))$. We also define
the shorthand $A(\epsi)=\Or_0( \epsi^n)$, where the subscript
expresses that a family $A(\epsi)\in \B(\Hi)$ is $\Or( \epsi^n)$
in the norm of bounded operators. By $A(\epsi)=\Or_0(
\epsi^\infty)$ we mean that $A(\epsi)=\Or_0( \epsi^n)$ for any $n
\in \N$. The remaining notation is defined in Appendix A.

\begin{proposition}

Let $\{E_n\}_{n\in\mathcal{I}}$ be an isolated family of bands and let
Assumption (A$_1$) be satisfied. Then there exists an orthogonal
projection $\Pi^\epsi \in \B(\Ht)$ such that \be \big[ \, H^\epsi_{\rm Z}
, \,\Pi^\epsi \,\big] = \Or_0(\epsi^\infty) \label{HP commutator} \ee
and $\Pi^\epsi =\widehat{\pi }+\Or(\epsi^{\infty })$, where
$\widehat{\pi }$ is the Weyl quantization of a $\tau$-equivariant
semiclassical symbol
\[
\pi \asymp \sum_{j\geq 0}\varepsilon ^{j}\pi _{j}\quad
\mathrm{in} \quad S^{1}_{\tau }(\epsi, \B(\Hi_{\rm f}))\,,
\]
whose principal part $\pi _{0}(k,r)$ is the spectral projector of
$H_{0}(k,r)$ corresponding to the given isolated family of bands.
\label{Prop Invariant subspace}
\end{proposition}

\begin{proof}
We first construct $\pi$ on a formal symbol
 level.

\begin{lemma}\label{Lemma Moyal proj}
Let $w(k,r)=(1 + k^2)$. There exists a {\em unique} formal symbol
\[
\pi = \sum_{j=0}^\infty \epsi^j\pi_j \quad \in \,M^{1}_\tau(\epsi,
\B(\Hi_{\rm f}))\cap M^{w}_\tau(\epsi,\B(\Hi_{\rm f},\Do))
\]
such that $\pi_0(k,r) = P_{\mathcal{I}}\big(k-A(r)\big)$ and
\begin{enumerate}
\item $\pi\,\sharp  \,\pi = \pi$,
\item $\pi^* = \pi$,
\item $H\,\sharp  \,\pi - \pi\,\sharp  \, H = 0$.
\end{enumerate}
\end{lemma}

\begin{proof}

We construct the formal symbol $\pi$ locally in
phase space and obtain by uniqueness, which can be proved as in
\cite{PST1}, a globally defined formal symbol.

Fix a point $z_0=(k_0,r_0) \in \R^{2d}$. From the continuity of
the map $z\mapsto H(z)$ and the gap condition it follows that
there exists a neighborhood $\U_{z_0}$ of $z_0$ such that for
every $z\in\U_{z_0}$ the set $\{E_n(z)\}_{n\in\mathcal{I}}$ can be
enclosed by a positively-oriented circle $\Lambda(z_0)\subset \C$
independent of $z$ in such a way that $\Lambda(z_0)$ is symmetric
with respect to the real axis,
\be\label{5Cg}
{\rm
dist}\big(\Lambda(z_0), \sigma(H(z))\big) \geq \frac{1}{4}C_{\rm
g} \quad\mbox{for all} \quad z\in\U_{z_0}
\ee
and
\be\label{5Cr}
{\rm Radius}(\Lambda(z_0))\leq C_{\rm r}\,.
\ee
The constant
$C_{\rm g}$ appearing in (\ref{5Cg}) is the same as in Definition
\ref{Isodef} and the existence of a constant $ C_{\rm r}$
independent of $z_0$ such that (\ref{5Cr}) is satisfied follows
from the periodicity of $ \{E_n(z)\}_{n\in\mathcal{I}}$ and the
fact that $A$ and $\phi$ are bounded. Indeed, $\Lambda$ can be chosen
$\Gamma^*$-periodic, i.e.\ such that $\Lambda(k_0+\gamma^*,r_0)=
\Lambda(k_0,r_0)$ for all $\gamma^*\in \Gamma^*$.


Let us choose any $\zeta\in \Lambda(z_0)$ and restrict all the
following expressions to $z\in\U_{z_0}$. We will construct a
formal symbol $R(\zeta)$ with values in $\B(\Hi_{\rm f},\Do)$ ---
the local Moyal resolvent of $H$ --- such that
\be\label{5res}
(H-\zeta)\,\sharp  \,R(\zeta)= {\bf 1}_{\Hi_{\rm f}}
\quad\mbox{and}\quad R(\zeta)\,\sharp  \,(H-\zeta) = {\bf 1}_\Do
\quad\mbox{on}\,\,\U_{z_0}\,.
\ee
To this end let
\[
R_0(\zeta) = (H-\zeta)^{-1}\,,
\]
where  according to (\ref{5Cg})  $R_0(\zeta)(z)\in \B(\Hi_{\rm
f},\Do)$ for all $z\in\U_{z_0}$, and, using differentiability of
$H(z)$, $\partial^\alpha_z R_0(\zeta)(z)\in \B(\Hi_{\rm f},\Do)$
 for all $z\in\U_{z_0}$. By construction one has

\[
(H-\zeta)\,\sharp  \,R_0(\zeta) = {\bf 1}_{\Hi_{\rm f}} +
\Or_0(\epsi)\,,
\]
where the remainder is $\Or(\epsi)$ in the $\B(\Hi_{\rm f})$-norm.
We proceed by induction. Suppose that
\[
R^{(n)}(\zeta) = \sum_{j=0}^n \epsi^j R_j(\zeta)
\]
with $R_j(\zeta)(z) \in \B(\Hi_{\rm f},\Do)$ for all $z\in\U_{z_0}
$ satisfies the first equality in (\ref{5res}) up to
$\Or(\epsi^{n+1})$, i.e.
\be
\label{R H} (H-\zeta) \,\sharp  \,R^{(n)} (\zeta) = {\bf 1}_{\Hi_{\rm
f}} + \epsi^{n+1} E_{n+1}(\zeta) + \Or_0(\epsi^{n+2})\,,
\ee
 where
$E_{n+1}(\zeta)(z) \in \B(\Hi_{\rm f})$. By choosing
\be  \label{R n+1}
R_{n+1}(\zeta) =  - R_0(\zeta)\, E_{n+1}
\ee
we obtain that $R^{(n+1)}(\zeta) = R^{(n)}(\zeta) +
\epsi^{n+1}R_{n+1}(\zeta)$ takes values in $\B(\Hi_{\rm f},\Do)$
and satisfies  the first equality in (\ref{5res}) up to
$\Or(\epsi^{n+2})$. Hence the formal symbol $R(\zeta)=
\sum_{j=0}^\infty \epsi^j R_j(\zeta)$ constructed that way
satisfies  the first equality in (\ref{5res}) exactly. By the same
argument one shows that there exists a formal symbol
$\widetilde{R}(\zeta)$ with values in $\B(\Hi_{\rm f}, \Do)$ which
exactly satisfies the second equality in (\ref{5res}). By the
associativity of the Moyal product, they must agree:
\[
\widetilde{R}(\zeta) = \widetilde{R}(\zeta)\, \sharp
\,(H-\zeta)\,\sharp  \,R(\zeta) = R(\zeta) \qquad \mbox{on  } \U_{z_0}.
\]

  Equations  (\ref{5res}) imply that $R(\zeta)$ satisfies the
resolvent equation
\be\label{5reseq}
R(\zeta) - R(\zeta') = (\zeta-\zeta')\,R(\zeta)
\,\sharp  \,R(\zeta')\quad\mbox{on}\,\,\U_{z_0}
\ee
for any
$\zeta,\zeta'\in \Lambda(z_0)$.  From the resolvent equation it
follows as in \cite{PST1} that the
 $\B(\Hi_{\rm f},\Do)$-valued formal symbol $\pi = \sum_{j=0}^\infty
\epsi^j\pi_j$ defined through
\be\label{5pidef}
\pi_j(z) :=
\frac{\I}{2\pi}\oint_{\Lambda(z_0)} \D\zeta\,
R_j(\zeta,z)\quad\mbox{on}\,\,\U_{z_0}
\ee
satisfies (i) and (ii)
of Lemma \ref{Lemma Moyal proj}. As for (iii) a little bit of care
is required. Let $J:\Do \to\Hi_{\rm f}$ be the continuous
injection of $\Do$ into $\Hi_{\rm f}$. Using (\ref{5pidef}) and
(\ref{5reseq}) it follows that $\pi\, J\,\sharp  \,R(\zeta) =
R(\zeta)\,J\,\sharp  \,\pi$ for all $\zeta\in \Lambda(z_0)$.
Moyal-multiplying from left and from the right with $H-\zeta$ one
finds $H\,\sharp  \,\pi\, J= J\, \pi \,\sharp  \,H$ as operators in
$\B(\Do,\Hi_{\rm f})$. However, by construction $H\,\sharp  \,\pi$ takes
values in $\B(\Hi_{\rm f})$ and, by density of $\Do$, the same
must be true for $\pi\,\sharp  \,H$.

We are left to show that $\pi \in \,M^1_\tau(\epsi, \B(\Hi_{\rm
f}))\cap M_\tau^w(\epsi, \B(\Hi_{\rm f},\Do))$. To this end notice
that by construction $\pi$ inherits the $\tau$-equivariance of
$H$, i.e.\
\[
\pi_j (k-\gamma^*,q) = \tau(\gamma^*)\,\pi_j(k,q)\,
\tau(\gamma^*)^{-1}\,.
\]
From (\ref{5pidef}) and (\ref{5Cr}) we conclude that for each
$\alpha\in\N^{2d}$ and $j \in \N$ one has
 \be \label{5bound1}
\|(\partial^\alpha_z\pi_j)(z)\|\leq 2\pi C_{\rm
r}\,\sup_{\zeta\in\Lambda(z_0)}\| (\partial^\alpha_z
R_j)(\zeta,z)\|\,,
\ee
where $\|\cdot\|$ stands either for the
norm of $\B(\Hi_{\rm f})$ or for the norm of $\B(\Hi_{\rm
f},\Do)$. In order to show that  $\pi \in \,M^{1}_\tau(\epsi,
\B(\Hi_{\rm f}))$ it suffices to consider  $z =(k,r) \in
M^*\times\R^d$ since $\tau(\gamma^*)$ is unitary and thus the
$\B(\Hi_{\rm f})$-norm of $\pi$ is periodic. According to
(\ref{5bound1}) we must show that
\be \label{Rj goal} \|(\partial_z^\alpha R_j)(\zeta,z)
\|_{\B(\Hi_{\rm f})} \leq C_{\alpha j}\quad \forall \, z\in
\U_{z_0}, \, \zeta\in\Lambda(z_0)
\ee
with $C_{\alpha j}$
independent of $z_0 \in M^* \times \R^d $.

We prove (\ref{Rj goal}) by induction. Assume, by
induction hypothesis, that for any $j \leq n$ one has that
 \be
\label{Inductive hp} R_j(\zeta) \in S_{\tau}^1(  \B(\Hi_{\rm
f}))\cap S_{\tau}^w( \B(\Hi_{\rm f}, \Do))
\ee
uniformly in $\zeta$,
in the sense that the Fr\'echet semi-norms are bounded by
$\zeta$-independent constants. Then, according to Proposition
\ref{ABProSymbComp}, $E_{n+1}(\zeta)$, as defined by (\ref{R H}),
belongs to $S_{\tau}^{w^2}(  \B(\Hi_{\rm f}))$ uniformly in
$\zeta$. By $\tau$-equivariance, the norm of $E_{n+1}(\zeta)$ is
periodic and one concludes that $E_{n+1}(\zeta) \in S_{\tau}^1(
\B(\Hi_{\rm f}))$ uniformly in $\zeta$. It follows from (\ref{R
n+1}) that (\ref{Inductive hp}) is satisfied for $j = n+1$.

  We are left to show that (\ref{Inductive hp}) is fulfilled
for $j = 0$. We notice that according to (\ref{5Cg}) one has for
all $z\in\R^{2d}$
\[
\|R_0(\zeta)\|_{\B(\Hi_{\rm f})}= \| (H(z) -
\zeta)^{-1}\|_{\B(\Hi_{\rm f})}= \frac{1}{{\rm dist} (\zeta,
\sigma(H(z)))} \leq \frac{4}{C_{\rm g}}\,.
\]
By the chain rule,
\be  \|(\partial_z R_0)(\zeta,z) \|_{\B(\Hi_{\rm
f})} = \| \big( R_0(\zeta) (\partial_z H_0)  R_0(\zeta) \big)(z)
\|_{\B(\Hi_{\rm f})}\,.
\ee
Since $\partial_z H_0 \, R_0(\zeta) $ is
a $\tau$-equivariant $\B(\Hi_{\rm f})$-valued symbol, its norm is
periodic. Therefore it suffices to estimate its norm for $z \in M^* \times \R^d$, which yields
the required bound. For a general $\alpha \in \N^{2d}$,
the norm of $\partial_z^{\alpha} R_0(\zeta)$ can be bounded in a
similar way. This proves that $R_0(\zeta)$ belongs to $S_{\tau}^1(
\B(\Hi_{\rm f}))$ uniformly in $\zeta$.

  On the other hand
\begin{eqnarray*}
\|R_0(k,r)\|_{\B(\Hi_{\rm f},\Do)}&=& \| (1+\Lap_x)\,
R_0([k]+\gamma^*,r) \|_{\B(\Hi_{\rm f})} \\ &=&  \| (1+\Lap_x)\,
\tau(\gamma^*) R_0([k],r)
\tau^{-1}(\gamma^*) \|_{\B(\Hi_{\rm f})}\\
&\leq&  C \, \| (1 + {\gamma^*}^2) (1+\Lap_x)\, R_0([k],r)
\|_{\B(\Hi_{\rm f})} \\&\leq&  C' (1 + {\gamma^*}^2)  \leq  2C' (1 + k^2)\,,
\end{eqnarray*}
where we used the fact that $ \| (1+\Lap_x) R_0(z) \|_{\B(\Hi_{\rm
f})} $ is bounded for $ z \in M^* \times \R^d$. The previous
estimate and the fact that $\partial_z H_0 \,\, R_0(\zeta)\in
S_{\tau}^1(  \B(\Hi_{\rm f}))$ yield
\begin{eqnarray*}
\|(\partial_z R_0)(\zeta,z) \|_{\B(\Hi_{\rm f}, \Do)}& = &\| \big(
R_0(\zeta) (\partial_z H_0) R_0(\zeta) \big)(z) \|_{\B(\Hi_{\rm
f}, \Do)}  \\&\leq& C (1 + k^2) \,.
\end{eqnarray*}
Higher order derivatives, are bounded by the same argument,
yielding that $R_0(\zeta)$ belongs to $S_{\tau}^w( \B(\Hi_{\rm f},
\Do))$ uniformly in $\zeta$. This concludes the induction
argument.

From the previous argument it follows moreover that
\be \label{Rj goal2}
\|(\partial_z^\alpha R_j)(\zeta,z) \|_{\B(\Hi_{\rm f},
\Do)} \leq C_{\alpha j} \ w(z)    \quad \forall \ z \in \U_{z_0},
\ \zeta\in\Lambda(z_0)
\ee
with $C_{\alpha j}$ independent of $z_0
\in \R^{2d}$. By (\ref{5bound1}), this implies   $\pi \in
\,M^{w}_\tau(\epsi, \B(\Hi_{\rm f}, \Do))$ and concludes the
proof.
 \end{proof}

\textit{Proof of Proposition \ref{Prop Invariant subspace}.}
   From the projector constructed in Lemma \ref{Lemma Moyal
proj} one obtains, by resummation, a semiclassical symbol $\pi \in
S^{1}_\tau(\epsi,\Hi_{\rm f})$ whose asymptotic expansion is given
by $\sum_{j\geq 0}\varepsilon ^{j}\pi _{j}$. Then  according to
Proposition \ref{ABThCaldVailltorus} Weyl quantization  yields a
bounded operator $\widehat{\pi }\in \B(\Ht)$, which is
approximately  a
 projector  in the sense that
\[
\widehat{\pi }^{2}=\widehat{\pi }+\Or_0(\epsi^{\infty })\, \, \,
\mathrm{and\, \, \, }\widehat{\pi }^{*}=\widehat{\pi }\,.
\]

We notice that   Proposition \ref{ABProSymbComp} implies that $ H
\, \widetilde{\sharp  } \, \pi \in S_{\tau}^{w^2}( \epsi,\B(\Hi_{\rm
f}))$. But $\tau$-equivariance implies that the norm is periodic
and then $H \, \widetilde{\sharp  } \, \pi$ belongs indeed to
$S_{\tau}^1(\epsi, \B(\Hi_{\rm f}))$. Then $ \pi \, \widetilde{\sharp  }
\, H = \big( H \, \widetilde{\sharp  } \, \pi \big)^{*}$ belongs to the
same class, so that $[H, \pi]_{\widetilde{\sharp  }} \in
S_{\tau}^1(\epsi,\B(\Hi_{\rm f}))$. This \textit{a priori}
information on the symbol class, together with Lemma \ref{Lemma
Moyal proj}.(iii), assures that
\be
\lbrack \widehat{H},\widehat{\pi } \rbrack =\Or_0(\epsi^{\infty })
\label{pi commutator}
\ee
with the remainder bounded in the $\B(\Ht)$-norm.

In order to get a true projector, we proceed as in \cite{NeSo}.
For $\epsi$ small enough, let
\be \label{Riesz2}
\Pi^\epsi :=\frac{\I}{2\pi }\int_{|\zeta -1|=\frac{1}{2}}\D\zeta
\,(\widehat{\pi }-\zeta )^{-1}\,.
\ee
Then it follows   that $ \Pi^{\epsi\,2}=\Pi^\epsi $, $\Pi^\epsi
=\widehat{\pi }+\Or_0(\epsi^{\infty })$ and
\[
\| \,\lbrack \widehat{H},\Pi^\epsi ]\,\| _{\B(\Ht)}\leq C  \|\,
[\widehat{H},\widehat{\pi }]\,\| _{ \B(\Ht)}=\Or (\epsi^{\infty
})\,.
\]
 \end{proof}

\subsection{The intertwining unitaries}

After having determined the decoupled subspace associated with an
isolated family of Bloch  bands, we aim at an effective description of the intraband dynamics, i.e.\
the dynamics inside this subspace. In order to get a workable
formulation of the effective dynamics, it is convenient to map the
decoupled subspace to a simpler reference space. The natural reference Hilbert space for the effective
dynamics is $\K := L^2(\T^{d *}) \otimes \C^\ell $, where $\ell:= {\rm
dim} P_{\mathcal{I}}(k)$ and $\T^{d *}$ is $M^*$ with periodic boundary conditions. Notation will be simpler in the
following, if we think of the fibre $\C^\ell$ as a subspace of
$\Hi_{\rm f}$. In order to construct such a unitary mapping, we
reformulate
 Assumption (A$_2$).\\

 \noindent {\bf Assumption A$_2'$.}\quad{\em
 Let
 $\{E_n(k)\}_{n\in\mathcal I}$ be an isolated family of bands and
  let $\pi_\mathrm{r} \in \B(\Hi_{\rm f})$ be an orthogonal projector with  ${\rm dim}\pi_{\rm
  r}=\ell$.
  There is a
unitary-operator-valued map $ u_{0}: \R^{2d} \rightarrow
\U(\Hi_{\rm f}) $ so that
\be
 u_{0}(k,r) \, \pi_{0}(k,r) \, u_{0}^{*}(k,r)=
\pi_{\mathrm{r}} \label{intertwining}
\ee
for any $(k,r) \in
\R^{2d}$,
\be
u_{0}(k+\gamma^*, r)=u_{0}(k,r) \tau (\gamma^* )^{-1} \,,
\label{T-covariance right}
\ee
and $u_{0}$ belongs to $S^1(  \B(\Hi_{\rm f}))$. }\\

  Clearly,
\be
u_{0}^{*}(k + \gamma^* ,r)=\tau (\gamma^* )u_{0}^{*}(k,r).
\label{T-covariance left}
\ee
  An operator-valued symbol satisfying (\ref{T-covariance
left}) (resp.\ (\ref{T-covariance right})) is called left $\tau
$-covariant (resp.\ right $\tau $-covariant).

The equivalence of (A$_2$) and (A$'_2$) can be seen as follows.
According to Assumption (A$_2$), there exists an orthonormal basis
$\left\{ \psi _{j}(k) \right\}_{j=1}^{\ell}$ of ${\rm
Ran}P_{\mathcal{I}}(k)$ which is smooth and $\tau$-equivariant
with respect to $k$. Let $\pi_{\rm r}:= \pi_0(k_0, r_0)$ for any
fixed point $(k_0, r_0)$. By the gap condition, ${\rm dim}\pi_{\rm
r} = {\rm dim}P_{\mathcal{I}}(k) $. Then for
any orthonormal basis $\left\{ \chi _{j}\right\}^{\ell}_{j=1}$ for
${\rm Ran}\pi_{\rm r}$, the formula
\be
\label{u0 explicit} \widetilde{u}_{0}(k,r):=\sum_{j = 1}^{\ell}
\left| \chi_{j}\right\rangle \left\langle \psi_{j}(k -A(r))\right|
\ee
 defines a partial isometry which can be extended to a unitary
operator $u_0(k,r) \in \U(\Hi_{\rm f})$. The fact that $\left\{
\psi _{j}(k) \right\}_{j=1}^{\ell}$ spans ${\rm
Ran}P_{\mathcal{I}}(k)$ implies (\ref{intertwining}), and the
$\tau$-equivariance of $\psi_j(k)$ reflects in (\ref{T-covariance
right}).

Viceversa, given $u_{0}$ fulfilling Assumption (A$'_2$), one can
check that the formula
\[
\psi_j(k - A(r)) := u_0^*(k,r) \chi_j,
\]
with $\left\{ \chi _{j}\right\} _{j =1}^{\ell}$ spanning ${\rm Ran}
\pi_{\rm r}$, defines an orthonormal basis for ${\rm Ran}
P_{\mathcal{I}}(k)$ which satisfies Assumption (A$_2$).

After these remarks recall that the goal of this section is to
construct a unitary operator which allow us to map the intraband
dynamics from ${\rm Ran}\Pi^\epsi$ to an $\epsi$-independent reference
space $\K \subset \Href$. Since all the twisting of $\Ht$ has been
absorbed in the $\tau$-equivariant basis
$\left\{ \psi_{j}\right\}_{j =1}^{\ell}$, or equivalently in $u_0$, the space
$\Href$ can be chosen to be a space of \emph{periodic}
vector-valued functions, i.e.
\[
\Href := L^2_{\tau \equiv {\bf 1}}(\R^d, \Hi_{\rm f}) \cong
L^2(\T^{d *}, \Hi_{\rm f}).
\]
 We introduce the orthogonal projector $\Pi_{\rm
r} := \hat{\pi}_{\rm r} \in \B(\Href)$ since the effective
intraband dynamics can be described in
\[
\K := {\rm Ran}\Pi_{\rm r} \cong L^2_{\tau \equiv {\bf 1}}(\R^d,
\C^\ell) \cong L^2(\T^{d *}, \C^\ell)
\]
as it will become apparent later on. Recall that
$\ell = {\rm dim}P_{\mathcal{I}}(k) = {\rm dim} \pi_{\rm r}$.

\begin{proposition}
\label{Prop unitaries} Let $\{E_n\}_{n\in\mathcal{I}}$ be an
isolated family of bands  and let
Assumptions \emph{(A$_1$)} and \emph{(A$'_2$)} be satisfied. Then
there exists a unitary operator $U^\epsi:$ $  \Ht \rightarrow \Href$ \
such that
\be
U^\epsi\, \Pi^\epsi \, U^{\epsi\,*}=\Pi _{\mathrm{r}}  \label{Intertwines}
\ee
and $U^\epsi=\hat{u}+\Or_0(\epsi^{\infty })$, where $u\asymp \sum_{j\geq
0}\varepsilon ^{j}u_{j}$ belong to $S^{1}(\epsi ,\B(\Hi_{\rm
f}))$, is right $\tau$-covariant at any order and has principal
symbol $u_{0}$.
\end{proposition}

\begin{proof}
By using the same method as in Lemma~3.3 in \cite{PST1}, one constructs first the formal symbol 
$\sum_{j\geq
0}\varepsilon ^{j}u_{j}$.
   Since
$u_{0}$ is right $\tau $-covariant, one proves by induction that
the same holds true for any $u_{j}$. Indeed, by referring to the
notation in \cite{PST1}, one has that
\[
u_{n+1}=(a_{n+1}+b_{n+1})u_{0}
\]
with $a_{n+1}=-\frac{1}{2}A_{n+1}$ and
 $b_{n+1}=[\pi_{r},B_{n+1}]$. From the defining equation
\[
u^{(n) }\ \sharp  \ u^{(n)*}-1=\varepsilon
^{n+1}A_{n+1}+\mathcal{O}(\varepsilon ^{n+2})
\]
and the induction hypothesis, it follows that $A_{n+1}$ is a
periodic symbol. Then $w^{(n)}:=u^{(n)}+\varepsilon
^{n+1}a_{n+1}u_{0}$ is right $\tau $-covariant. Then the defining
equation
\[
w^{(n) }\ \sharp  \ \pi \ \sharp  \ w^{(n)*}-\pi _{\mathrm{r }}=\varepsilon
^{n+1}B_{n+1}+\mathcal{O}(\varepsilon ^{n+2})
\]
shows that $B_{n+1}$ is a  periodic symbol, and so is $b_{n+1}$.
Hence $u_{j}$ is right $\tau $-covariant, and there exists a
semiclassical symbol $u\asymp \sum_{j}\varepsilon ^{j}u_{j}$ so
that $u\in S^{1}(\epsi,\B(\Hi_{\rm f}))$.

   One notices that right
$\tau$-covariance is nothing but a special case of $(\tau_1,
\tau_2)$-equivariance, for $\tau_2 \equiv {\bf 1}$ and $\tau_1 =
\tau $. Thus it follows from Proposition \ref{ABThCaldVailltorus}
that the Weyl quantization of $u$ is a bounded operator
$\widehat{u}\in \B( \Ht,\Href) $\ such that:

\begin{enumerate}
\item  $\widehat{u}\,\widehat{u}^{*}={\bf 1}_{\Href}+\Or_0(\epsi^{\infty
})$\quad and \quad $\widehat{u}^{*}\widehat{u}= {\bf
1}_{\Ht}+\Or_0(\epsi ^{\infty })$,
\item  $\widehat{u}\,\Pi^\epsi \, \widehat{u}^{*}=\Pi_{\rm r}+\Or_0(\epsi ^{\infty })$.
\end{enumerate}

Finally we modify $\widehat{u}$ as in \cite{PST1} by an
$\Or_0(\epsi^{\infty})$-term in order to get the   unitary
operator $U^\epsi\in \U(\Ht, \Href)$.     \end{proof}

\subsection{The effective Hamiltonian}

The  final step in space-adiabatic perturbation theory is to
define and compute the effective Hamiltonian for the intraband
dynamics and to compute its lower order terms. This is done, in
principle, by projecting the full Hamiltonian $H^\epsi_{\rm Z}$
to the decoupled subspace and afterwards rotating to the reference
space.

\begin{proposition}
\label{Prop Heff} Let $\{E_n\}_{n\in\mathcal{I}}$ be an isolated
family of bands and let Assumptions $(A_1)$ and $(A_2)$ be satisfied.
Let $h$ be a resummation in $S^{1}_{\tau \equiv {\bf
1}}(\epsi,\B(\Hi_{\rm f}))$ of the formal symbol \be h=u  \,
\sharp   \, \pi  \, \sharp \, H \, \sharp   \, \pi \, \sharp \,
u^{*} \,\in\, M^{1}_{\tau \equiv {\bf 1}}(\epsi,\B(\Hi_{\rm f}))
\,. \label{heffective2} \ee
 Then $\widehat h  \in \B(\Href)$,
$ [  \widehat h,\Pi_{\rm r} ] =0$ and
\be
\big( \E^{-\I  H^\epsi_{\rm Z} t} -U^{\epsi\,*}\, \E^{-\I \widehat h
t}\, U^\epsi \big) \Pi^\epsi  =\Or_0(\epsi^{\infty }(1+|t|))\,. \label{Heff
unitaries}
\ee
\end{proposition}

\begin{remark}
The definition of the effective Hamiltonian is not entirely unique in the sense
that any $H_{\rm eff}$ satisfying (\ref{Heff unitaries}) would
serve as well as an effective Hamiltonian. However, the asymptotic expansion of $H_{\rm eff}$
is unique and therefore it is most convenient to define the effective Hamiltonian
through (\ref{heffective2}).\er
\end{remark}

\begin{proof}
In the proof we denote $\Ham$ as $\widehat{H}$ to emphasize the
fact that it is the Weyl quantization of $H \in S_{\tau}^w(\epsi,
\B(\Do, \Hi_{\rm f}))$.

First note that (\ref{heffective2}) follows from the following
facts: according to Lemma  \ref{Lemma Moyal proj} and Proposition
\ref{ABProSymbComp} we have that
\[
\pi\,\sharp  \, H \, \sharp  \, \pi \in M_{\tau}^{w^2} (\epsi, \B(  \Hi_{\rm
f})) = M_{\tau}^{1} (\epsi, \B(  \Hi_{\rm f}))\,,
\]
where we used that $\tau$ is a unitary representation.
With Proposition \ref{Prop unitaries} it follows that $h\in
M_{\tau\equiv 1}^{1} (\epsi, \B(  \Hi_{\rm f}))$. Therefore
$\widehat h \in \B(\Href)$ follows from Proposition
\ref{ABThCaldVailltorus}, while $[ \widehat h,\Pi_{\rm r} ] =0$ is
satisfied by construction.

It remains to check (\ref{Heff unitaries}):
\begin{eqnarray*}
\big(   \E^{-\I  \widehat{H} t}- U^{\epsi\,*}\, \E^{-\I \widehat h t}\,U^\epsi
 \big) \Pi^\epsi &=& \big(   \E^{-\I  \widehat{H} t}-  \E^{-\I U^{\epsi\,*} \,\widehat h \,U^\epsi t}
 \big) \widehat \pi +\Or_0(\epsi^\infty) \\
 &=&
 \big(   \E^{-\I  \widehat \pi \widehat{H}\widehat \pi t}-  \E^{-\I U^{\epsi\,*} \,\widehat h \,U^\epsi t}
 \big) \widehat \pi +\Or_0(\epsi^\infty)\\&=& \Or(\epsi^\infty(1+|t|))\,,
\end{eqnarray*}
where the last equality follows from the usual Duhammel argument and the
fact that the difference of the generators is
$\Or_0(\epsi^\infty)$ in the norm of bounded operators by
construction.
 \end{proof} 

Since $[ \widehat{h} ,\Pi_{\rm r }] =0$, the effective Hamiltonian
will be regarded, without distinctions in notation, either as an
element of $\B(\Href)$ or as an element of $\B(\K)$.

We  compute the principal and the subprincipal symbol of $\widehat
h$ for the special but most relevant case of an isolated
eigenvalue, eventually $\ell$-fold degenerate, i.e. $E_n(k)\equiv
E(k)$ for every $n \in \mathcal{I}, \, |\mathcal{I}|= \ell $.
Recall that in this special case Assumption ($A_2$) is equivalent
to the existence of an orthonormal system of smooth and
$\tau$-equivariant Bloch functions corresponding  to the
eigenvalue $E(k)$. If $\ell=1$ then Assumption (A$_2$) is always
satisfied. The part of $u_0$ intertwining $\pi_0$ and $\pi_{\rm
r}$ is given by equation (\ref{u0 explicit}) where $\psi_j(k)$ are
now Bloch functions, i.e. eigenvectors of $H_{\rm per}(k)$ with
eigenvalue $E(k)$.

\begin{proof}[Proof of Corollary \ref{MainCor}]
In the following  $h$ is identified with $ \pi_{\rm
r}h  \pi_{\rm r}$ and regarded as a $\B(\C^\ell)$-valued symbol. We
consider the matrix elements
\[
h(k,r)_{\alpha
\beta} := \langle \chi_{\alpha}, h (k,r)
\chi_{\beta} \rangle
\]
for $\alpha, \beta \in \{1, \ldots, \ell
\}$, where we recall that $\chi_\alpha = u_0(k,r) \psi_\alpha(k-A(r))$.
 Equation (\ref{h0 specialbis}) follows immediately from the fact
that $h_0 = u_0 \, H_0 \, u_0^{*}$ and that $\psi_{\alpha}$ are
Bloch functions.  As for $h_1$, we use the general formula  
 of \cite{PST1}, which reads, transcribed to the present
setting, as
\begin{eqnarray}\label{h1 specialgen}
h_{1\,\alpha \beta }(k,r) &=& -\I\,\big\langle \psi_\alpha(\widetilde k), \,\{E(\widetilde k) +\phi(r) , \psi_\beta(\widetilde k)
\}\big\rangle \nonumber\\&&-\,\textstyle{\frac{\I}{2}}\big\langle \psi_\alpha(\widetilde k) ,\{(H_{\rm per}(\widetilde k)-E(\widetilde k)
),\psi_\beta(\widetilde k) \}\big\rangle \,.
\end{eqnarray}
Here $\{A,\ph\}= \nabla_r A\cdot \nabla_k \ph - \nabla_k A\cdot \nabla_r \ph$ are the Poisson brackets
for an operator-valued function $A(k,r)$ acting on a vector-valued function $\ph(k,r)$.
We need to evaluate (\ref{h1 specialgen}).  Inserting (\ref{u0 explicit}) and
performing a straightforward computation  the first term in
(\ref{h1 specialgen}) gives the first term in (\ref{h1 special})
while the second term contributes to the $\alpha
\beta$ matrix element with
\[
\frac{\I}{2} \sum_{j,l=1}^{d} \big(\partial_j A_l  - \partial_l
A_j \big)(r) \,\big\langle \psi_{\alpha}(\widetilde k),
\partial_l (H_{\rm per} - E)(\widetilde k) \
\partial_j \psi_{\beta} (\widetilde k) \big\rangle_{\Hi_{\rm f}}\,.
\]
The
derivative on $(H_{\rm per} - E)$ can be moved to the first
argument of the inner product by noticing that
\[
0 = \nabla
\big\langle \psi_{\alpha}, (H_{\rm per} - E) \phi \big\rangle =
\big\langle \nabla \psi_{\alpha}, (H_{\rm per} - E) \phi
\big\rangle + \big\langle \psi_{\alpha}, \nabla (H_{\rm per} - E)
\phi \big\rangle
\]
since $\psi_{\alpha}$ is in the kernel of
$(H_{\rm per} - E)$.  Finally the imaginary part of
\[
\frac{\I}{2} \sum_{j,l=1}^{d} \big(\partial_j A_l  - \partial_l
A_j \big)(r) \,\big\langle\partial_l  \psi_{\alpha}(\widetilde k),
\,(H_{\rm per} - E)(\widetilde k) \
\partial_j \psi_{\beta} (\widetilde k) \big\rangle_{\Hi_{\rm f}}
\]
vanishes, as can be seen  by direct computation, concluding the proof.
 \end{proof}

\section{Semiclassical dynamics  for   Bloch electrons}

We have now at our disposal the tools to establish the link 
between the Schr\"odin\-ger equation (\ref{basic_dynamics}) and the corrected semiclassical
equations of motion (\ref{Semi1}). To this end we specialize to the case of a non-degenerate
Bloch band $E_n$. The phase space for (\ref{Semi1}) is
$\R^d\times\R^d$, since we use the extended zone scheme, and we denote by $\Phi^t_\epsi$
the corresponding solution flow. Since the effective Hamiltonian is written in canonical 
variables, it is necessary to switch in (\ref{Semi1}) to $(r,k)$ with $k = \kappa+A(r)$.
In the new coordinates the solution flow is denoted by
$\overline\Phi^t_\epsi$ and 
 \[
\overline \Phi^t_{\epsi} (r,k) = \Big( \Phi^t_{\epsi
\,r}\big(r,k-A(r)\big),\,\Phi^t_{\epsi
\,\kappa}\big(r,k-A(r)\big)+A(r)\Big)\,.
\]
Let us consider any admissible semiclassical observable $\widehat a = a(\epsi x,-\I\nabla_x)$
acting on the ``physical'' Hilbert space $L^2(\R^d,\D x)$.
Its symbol is transported by $\overline\Phi^t_\epsi$ to $a\circ\overline\Phi^t_\epsi$
with Weyl quantization $\widehat{a\circ\overline\Phi^t_\epsi}$.
On the other hand the operator $\widehat a$ is transported by the Heisenberg equation as
$\E^{\I H^\epsi t/\epsi}\widehat a \E^{-\I H^\epsi t/\epsi}$. Our assertion is that on the subspace $\Pi_n^\epsi L^2(\R^d) :=
U^{\epsi\,*}\Pi^\epsi \Ht$, $\Pi^\epsi$ and $U^\epsi$ as constructed in the previous section, 
these two operators are uniformly close to order $\epsi^2$.

\begin{theorem} \label{EgCor}

Let $E_n$ be an isolated, non-degenerate Bloch band, see
Definition~\ref{Isodef}, and let the potentials satisfy Assumption
(A$_1$). Let $a\in C^\infty_{\rm b}(\R^{2d})$ be
$\Gamma^*$-periodic in the second argument, i.e.\ $a(r,k+\gamma^*)
= a(r,k)$ for all $\gamma^*\in\Gamma^*$, and $\widehat a = a(
\epsi x,-\I\nabla_x)$ be its Weyl quantization. Then for each
finite time-interval $I\subset \R$ there is a constant $C<\infty$
such that for $t\in I$
\[
 \left\|\, \Pi^\epsi_n\, \left(\, \E^{\I H^\epsi t/\epsi} \, \widehat a\,\,  \E^{-\I
  H^\epsi t/\epsi}\,-\, \widehat{ a\circ \overline \Phi^{t}_{\epsi} }\, \right)\,\Pi^\epsi_n \,\right\|_{\B(L^2(\R^d))}
  \leq \epsi^2\,C\,.
\]
In particular, for $\psi_0 \in \Pi^\epsi_n \Hi$   we have that
\be\label{Wig}
\big|\,\big\langle \psi_0, \,\E^{\I H^\epsi t/\epsi} \, \widehat
a\,\,  \E^{-\I
  H^\epsi t/\epsi}\,\psi_0\big\rangle - \big\langle \psi_0,\,\widehat{ a\circ \overline \Phi^{t}_{\epsi} }
  \,\psi_0\big\rangle \,\big| \leq \epsi^2\,C\,\|\psi_0\|^2\,.
\ee
\end{theorem}
Theorem~\ref{EgCor} is an Egorov-type theorem, see \cite{Ro}. An unconventional feature
is that the first order corrections are treated  by
considering an $\epsi$-dependent Hamiltonian flow instead of
having a separate dynamics for the subprincipal symbol of an
observable.

By exploiting the relation between Weyl-quantized  operators 
 and Wigner transforms, one can easily translate  (\ref{Wig}) to
 the language of Wigner functions. For a detailed discussion on how
 Theorem~\ref{EgCor} relates to alternative approaches to the semiclassical
 limit in perturbed periodic potentials we refer the reader to \cite{Te2}.

To prove Theorem~\ref{EgCor}, our strategy is to first establish a corresponding Egorov theorem
in the reference space and then
to pull back to $L^2(\R^d,\D x)$.

\begin{proposition} \label{Egorov}
Let $E$ be an isolated non-degenerate Bloch band and let $\widehat
h$ be the effective Hamiltonian constructed in Theorem \ref{Th
main}, which acts on the reference space $\K = L^2_{\tau \equiv
1}(\R^d)$ of\, $\Gamma^*$-periodic $L^2_{\rm loc}$-functions. Let
$\widetilde \Phi^t: \R^{2d}\to \R^{2d}$ be the Hamiltonian flow
generated by the   Hamiltonian function
\[
h_{\rm cl}(k,r) = h_0(k,r) + \epsi h_1(k,r)\,.
\]
Then for any semiclassical observable $\widehat a =
a_0(k,\I\epsi\nabla_k) + \epsi a_1(k,\I\epsi\nabla_k)$ with $a\in
S^{1}(\epsi,\C)$ we have that \be \big\|\, \E^{\I\widehat h
t/\epsi}\,\widehat a\,\E^{-\I\widehat h t/\epsi}\,-\, \widehat{ a
\circ \widetilde \Phi^t  }\,\big\| \leq C_T  \epsi^2 \ee uniformly
for any finite interval in time $[-T,T]$.
\end{proposition}
\begin{proof}
Since the Hamiltonian function is bounded with bounded derivatives,
it follows immediately that $a \circ \widetilde \Phi^t  \in S^{1}(\epsi)$
and that $\frac{\D}{\D t} \,( a \circ \widetilde \Phi^t ) \in S^{1}(\epsi)$.
Therefore the proof is just the standard computation
\begin{eqnarray*}\lefteqn{
\E^{\I\widehat h t/\epsi}\,\widehat a\,\E^{-\I\widehat h
t/\epsi}\,-\, \widehat{ a \circ \widetilde \Phi^t  }    =\int_{0}^{t}\,\D t'\,\frac{\D}{\D
t'}\left( \E^{\I \widehat{h}t'/\epsi }\,  \big( \widehat{ a \circ \widetilde \Phi^{t-t'}  }  \big)   \,\E^{-\I\widehat{h}t'/\epsi
}\right) } \\
&=&\int_{0}^{t}\,\D t'\,\E^{\I\widehat{h}t'/\epsi }\,
\left( \frac{\I}{\epsi }\left[ \,\widehat{h},
  \big( \widehat{ a \circ \widetilde \Phi^{t-t'}  }  \big) \,\right]
 -
 \Big(  \textstyle{\frac{\D}{\D t'}} \,( a \circ \widetilde \Phi^{t-t'})\Big)^{\widehat{\,\,}}\,  \right)
\E^{-\I\widehat{h}t'/\epsi }\,,
\end{eqnarray*}
together with the fact that the integrand is $\Or(\epsi^2)$ in the norm of bounded
operators, since
by construction
\[
 \frac{\D}{\D t'} \,( a \circ \widetilde \Phi^{t-t'} ) =
 \big\{ \, h_{\rm cl},\,a \circ \widetilde \Phi^{t-t'} \big\}
\]
and, computing the expansion of the Moyal product,
\[
\frac{\I}{\epsi }
\left[ \,h ,
     a \circ \widetilde \Phi^{t-t'}  \,\right]_{\widetilde \sharp  }=\big\{ \, h_{\rm cl},\,a \circ \widetilde \Phi^{t-t'}
     \big\}
     +\Or(\epsi^2)\,.
\] 
{} \end{proof} 

In order to obtain the Egorov theorem for the physical
observables, we need to undo the transform to the reference space
and the Zak transform. We start with the simpler observation on
how the Zak transform maps semiclassical observables.

\begin{proposition}\label{PerObsProp}
 Let $a\in S^1(\epsi,\C)$ be
  $\Gamma^*$-periodic, i.e.\ $a(r,k+\gamma^*) = a(r,k)$ for all
  $\gamma^*\in\Gamma^*$.
Let $b(k,r) = a(r,k)$ then $b\in S^1_\tau(\epsi,\C )$
and
\[
 \widehat  a = \U^*\,\widehat b\,\U\,,
\]
where  the Weyl quantization   is in the sense of\, $\widehat
a= a(\epsi x, -\I\nabla_x)$ acting on $L^2(\R^d)$ and $\widehat b = b(k,\epsi\I\nabla_k)$ acting
on $\Ht$.
\end{proposition}

\begin{remark}
 An analogous statement \emph{cannot} be true for   general
operator-valued $\tau$-equivariant symbols. For example, the
symbol $b(k,r):= H_{\rm per}(k - A(r))$ is $\tau$-equivariant and in
particular a semiclassical observable. However, the corresponding
operator in the original representation is
\[
\U^{*}\,\widehat b \,\U = - \frac{1}{2} \big(-\I \nabla_x -
A(\epsi x)\big)^2 + V_{\Gamma}(x)
\]
which cannot be written as a $\epsi$-pseudodifferential operator
with scalar symbol. \er
\end{remark}

\begin{proof} We give the proof for $a(\cdot,k)\in\Sch(\R^d)$. The
general result follows from standard density arguments, see \
\cite{DiSj}. For $\psi\in \Sch(\R^d)$ we have according to
(\ref{ABKer2}) the explicit formula \be \label{a quantized}
\big(a(\epsi x, -\I  \nabla_x) \psi \big)(x)=
\frac{1}{(2\pi)^{d/2}}\, \sum_{\gamma\in\Gamma}\int_{\R^{d}}
\D\eta   \big( \mathcal{F}a \big) (\eta, \gamma) \ \E^{\I \epsi
(\eta \cdot \gamma)/2} \E^{\I \epsi \eta \cdot x} \psi(x +
\gamma)\,. \ee On the other hand for $(\U\psi)(k,r)=: \ph(k,r)$ by
definition it holds that \be \label{B quantized} \big(b(k, \I
\epsi \nabla_k) \ph \big)(k,r)= \sum_{\gamma \in \Gamma}
\int_{\R^{d}} \D\eta \big( \F b\big) (\gamma, \eta) \ \E^{- \I
\epsi (\eta \cdot \gamma)/2} \E^{\I  \gamma \cdot k} \ph(k - \epsi
\eta, r)\,. \ee The assumptions on $a$ and $\psi$ guarantee that
all the integrals and sums in the following expressions are
absolutely convergent and thus that interchanges in the order of
integration are justified by Fubini's theorem.

We compute the inverse Zak transform of (\ref{B quantized}) using
(\ref{BFinv}),
\begin{eqnarray} \lefteqn{
 \big( \U^{-1} \widehat b\, \ph \big)(x)=} \label{Last observ} \\
&=& \sum_{\gamma \in \Gamma} \int_{B} \D k  \int_{\R^{d}} \D\eta\,
\big( \mathcal{F}b \big) (\gamma, \eta) \ \E^{\I k \cdot x} \E^{-
\I \epsi (\eta \cdot \gamma)/2} \E^{\I  \gamma \cdot k} \ph(k -
\epsi
\eta, [x]) \nonumber \\
&=& \sum_{\gamma \in \Gamma} \int_{\R^{d}} \D\eta \,\big(
\mathcal{F}b \big) (\gamma, \eta) \E^{\I \epsi (\eta \cdot
\gamma)/2} \E^{\I \epsi \eta \cdot x} \int_{M^{*}} \D k \, \E^{\I
(k - \epsi \eta) \cdot (x + \gamma)}  \ph(k - \epsi \eta,
[x])\,.\nonumber
\end{eqnarray}
The $\tau$-equivariance of $\ph$ implies that the function
$f(k,y):= \E^{\I k \cdot y} \ph(k, [y]) $ is exactly periodic in
the first variable. Then the integral in $\D k$ can be shifted by
an arbitrary amount, so that
\[
\int_{M^{*}} \D k \, \E^{\I (k - \epsi \eta) \cdot (x + \gamma)}
\ph(k - \epsi \eta, [x]) =  \int_{M^{*}} \D k \, \E^{\I k \cdot (x
+ \gamma)} \ph(k, [x + \gamma]) = \psi(x + \gamma)\,.
\]
Inserting this expression in the last line of (\ref{Last observ})
and comparing with (\ref{a quantized})   concludes the proof.
 \end{proof}

Before we arrive at the proof of Theorem \ref{EgCor}, one has to
study how the unitary map constructed in Section 3.2 maps
observables in the Zak representation to observables in the
reference representation.

\begin{proposition} \label{traPro}
Let $\widehat b = b_0(k,\epsi\I\nabla_k) +\epsi \,b_1(k,\epsi\I\nabla_k) $ with
symbol $b\in S^1(\epsi,\C)$ which is $\Gamma^*$-periodic in the first argument.
Let $U^\epsi: \Pi^\epsi\Ht \to \K$ be the unitary map constructed in Section 3.2.
Then
\[
U^\epsi\,\Pi^\epsi\,\widehat b\,\Pi^\epsi\,U^{\epsi\,*} = \widehat c + \Or(\epsi^2)\,,
\]
where $c(\epsi,k,r) =\big( b \circ T\big)(k,r)$ with
\[
T:  \R^{2d}\to\R^{2d}\,, \quad (k,r) \mapsto
\Big(k +\epsi\,  \A_m \big(k-A(r)\big)  \nabla A_m(r) ,\,r+\epsi \A
\big(k-A(r)\big)\Big)\,.
\]
Here and in the following, summation over indices appearing twice
is implicit.
\end{proposition}

\begin{proof}
In order to compute $c = u\,\sharp  \,\pi\,\sharp  \,b\,\sharp \,
\pi\,\sharp  \,u^*$, observe that, since $b$ is scalar-valued, the
principal symbol remains unchanged, i.e.\ $c_0 =
u_0\,\pi_0\,b_0\,\pi_0\,u_0^* = b_0$. For the subprincipal symbol
we use the general transformation formula (\ref{h1 specialgen})
obtained for the Hamiltonian, which applies to all operators whose
principal symbol commutes with $\pi_0$. In this case the
eigenvalue $E$ in (\ref{h1 specialgen}) must be replaced by the
corresponding principal symbol and a term for the subprincipal
symbol $b_1$ must be added.
Hence we find that
\begin{eqnarray*}
c_1(k,r) &= &- \I \,\big\langle \psi  (k-A(r)) ,\{ b_0(k,r) ,\psi
(k-A(r))
\}\big\rangle \\&& +\, \langle \psi  (k-A(r)) , b_1(k,r)\psi  (k-A(r))\rangle\\
&=& \partial_{k_n} b_0(k,r) \,  \I \,\big\langle \psi  (k-A(r)) , \partial_m \psi
(k-A(r))\big\rangle\,
\partial_n A_m(r)\\
&& +\,\partial_{r_n} b_0(k,r) \,  \I \,\big\langle \psi  (k-A(r)) , \partial_n \psi
(k-A(r))\big\rangle + b_1(k,r)\\&=&
 \partial_{k_n} b_0(k,r) \,\A_m   (k-A(r)) \,
\partial_n A_m(r)  \\&&+\,  \partial_{r_n} b_0(k,r)  \,\A_n  (k-A(r))
 + b_1(k,r)\,,
\end{eqnarray*}
where summation over indices appearing twice is implicit. Now a
comparison with the
Taylor expansion of  $\big( b \circ T \big)(k,r)$ in powers of $\epsi$  proves the
claim.
 \end{proof} 

We have now all the ingredients needed for the

\begin{proof}[Proof of Theorem \ref{EgCor}]
Let $a\in C^\infty_{\rm b}(\R^{2d})$ be
$\Gamma^*$-periodic in the second argument, then according to
Proposition \ref{PerObsProp} we have
\be\label{trafo3}
\Pi^\epsi_n\,  \E^{\I H^\epsi t/\epsi} \, \widehat a\,\,  \E^{-\I
  H^\epsi t/\epsi}\,\Pi^\epsi_n = \U^*\,\Pi^\epsi \,  \E^{\I H^\epsi_{\rm Z} t/\epsi} \, \widehat b\,\,  \E^{-\I
  H^\epsi_{\rm Z} t/\epsi}\,\Pi^\epsi\,\U
\ee
with $b(k,r) = a(r,k)$. With Theorem \ref{Th main} and Proposition \ref{traPro} we find that
\be\label{trafo1}
\Pi^\epsi \,  \E^{\I H^\epsi_{\rm Z} t/\epsi} \, \widehat b\,\,  \E^{-\I
  H^\epsi_{\rm Z} t/\epsi}\,\Pi^\epsi = U^{\epsi\,*}\, \E^{\I \widehat h t/\epsi} \, \widehat c\,\,  \E^{-\I
  \widehat h t/\epsi}\,U^\epsi + \Or(\epsi^2)\,,
\ee
where $c(\epsi,k,r) =\big( b \circ T\big)(k,r)$. Now we can
apply Proposition \ref{Egorov} to conclude that
\[
 \E^{\I \widehat h t/\epsi} \, \widehat c\,\,  \E^{-\I
  \widehat h t/\epsi} = \widehat{ \big( c\circ \widetilde
  \Phi^t\big) } + \Or(\epsi^2).
\]
Since, for $\epsi$ sufficiently small,  $T$ is a diffeomorphism,
one can write
\[
c\circ \widetilde \Phi^t  = c\circ T^{-1}\circ T\circ \widetilde \Phi^t \circ T^{-1}\circ T
=: c\circ T^{-1}\circ \overline \Phi^t \circ T  = b\circ \overline \Phi^t \circ T \,,
\]
where the flow $\overline \Phi^t_\epsi$ in the new coordinates
will be computed explicitly below. Inserting the results into
(\ref{trafo1}), one obtains
\begin{eqnarray*}
\Pi^\epsi \,  \E^{\I H^\epsi_{\rm Z} t/\epsi} \, \widehat b\,\,  \E^{-\I
  H^\epsi_{\rm Z} t/\epsi}\,\Pi^\epsi &= &U^{\epsi\,*}\, \widehat{\big( b\circ \overline \Phi^t \circ T\big)}
   \,U^\epsi + \Or(\epsi^2)\\
   &=& \Pi^\epsi \, \widehat{\big( b\circ \overline \Phi^t \big)}\,\Pi^\epsi  + \Or(\epsi^2)\,,
\end{eqnarray*}
where we used Proposition \ref{traPro} for the second equality.
Inserting into (\ref{trafo3}) we finally find that
\be
\Pi^\epsi_n\,  \E^{\I H^\epsi t/\epsi} \, \widehat a\,\,  \E^{-\I
  H^\epsi t/\epsi}\,\Pi^\epsi_n = \Pi^\epsi_n \, \widehat{\big( a\circ \overline \Phi^t  \big)}\,\Pi^\epsi_n
  + \Or(\epsi^2)\,,
\ee
where we did not make the exchange of the order of the arguments
in $a$ explicit.

Since the flow is determined only in approximation and only
through its vector field, we make use of the following lemma.

\begin{lemma} Let $\Phi_i:\R^{2d} \times \R\to \R^{2d}$ be
the flow associated with the vector field $v_i\in
C^\infty_{\rm b}(\R^{2d}, \R^{2d})$, $i=1,2$.
\begin{enumerate}
\item If for all $\alpha\in\N^{2d}$ there is a $c_\alpha<\infty$ such that
\[
\sup_{x\in\R^{2d}} |\,\partial^\alpha\,(v_1-v_2)(x) | \leq
c_\alpha \,\epsi^2\,,
\]
then  for each bounded interval $I\subset \R$ there are constants $C_{I,\alpha}<\infty$
such that
\be\label{lem1}
\sup_{t\in I, x\in\R^{2d}} |\,\partial^\alpha\,(\Phi^t_1-\Phi^t_2)(x) | \leq
C_{I,\alpha} \,\epsi^2\,.
\ee
\item Let $a\in S^1(\epsi,\C)$. If (\ref{lem1}) holds for the
flows $\Phi_1,\Phi_2$, then there is a constant $C<\infty$, such
that for all $t\in I$
\[
\big\|\, \widehat{a\circ\Phi^t_1} -
\widehat{a\circ\Phi^t_2}\,\big\|_{\B(L^2(\R^d))}\leq C\,\epsi^2\,.
\]
\end{enumerate}
\end{lemma}
\begin{proof}
Assertion (i) is a simple application of Gronwall's lemma.
Assertion (ii) follows from the fact that the norm of the
quantization of a symbol in $S^1$ is bounded by a constant times
the sup-norm of finitely many derivatives of the symbol, which are
$\Or(\epsi^2)$ according to (\ref{lem1}).
 \end{proof}

According to assertion (ii) of the lemma  it suffices to show that
\[
\overline \Phi^t_\epsi  (r,k) = \Big(
\Phi^t_{ \epsi\,r}(r,k-A(r)),\,\Phi^t_{\epsi\,\kappa}(r,k-A(r))+A(r)\Big)+\Or(\epsi^2)
\]
in the above sense, where $\Phi^t_n$ is the flow of (\ref{Semi1}).
And from assertion (i) we infer that it suffices to prove the
analogous properties on the level of the vector fields.

Through a subsequent change of coordinates we aim at computing the
vector field of $\Phi^t_\epsi$ up to an error of order $\Or(\epsi^2)$.
We start with the vector field of $\widetilde \Phi^t$.
The effective Hamiltonian on the reference space including first order
terms reads
\begin{eqnarray} \label{53Ham}
h( r,k)& =&   E(k - A(r)) + \phi(r)\\&& -\, \epsi\,\Big( F_{\rm
Lor}(r, \nabla E (k - A(r)))\cdot \A (k - A(r))  + B(r)\cdot M(k -
A(r))\Big)\,, \nonumber
\end{eqnarray}
with the Lorentz force
\[
F_{\rm Lor}(r, \nabla E(k - A(r))) = -\nabla\phi(r) + \nabla E(k - A(r)) \vp
B(r)\,.
\]
Componentwise, the canonical equations of motion are
\begin{eqnarray*}
\dot r_j & = & \partial_{k_j}  h(r,k)   =   \partial_{k_j} E(
k-A(r)) \\
&& - \epsi \,\partial_{k_j}\Big( F_{\rm Lor}(r,  k - A(r) )\cdot
\A (k - A(r))  +   B(r)\cdot M(k - A(r))\Big)\,,
\end{eqnarray*}
\begin{eqnarray*}\lefteqn{
\dot k_j  =  -\partial_{r_j} h(r,k) = -\partial_{j}\phi(r) +
\partial_{l } E(k-A(r))\partial_{j} A_l  (r) } \\
&& -\,\epsi\, \partial_{k_l }\Big( \A (k-A(r))\cdot F_{\rm Lor}(r,
k - A(r) ) + B(r)\cdot M(k-A(r))
\Big)\,\partial_{ j}A_l (r)\\
&& -\,\epsi \,\A_l (k-A(r)) \Big( \partial_j\partial_l \phi(r) -
\big(\nabla E(k-A(r))\vp \partial_j B(r)\big)_l  \Big)
\\&&+\, \epsi\,\partial_j
 B (r)\cdot M (k-A(r)) \,,
\end{eqnarray*}
with the convention to sum over repeated indices. Substituting
$\widetilde k = k- A(r)$ one obtains
\[
\dot r_j    =   \partial_{j} E( \widetilde k) - \epsi\,
\partial_{\widetilde k_j}\Big( F_{\rm Lor}(r, \widetilde k)\cdot \A (\widetilde k)
  + B(r)\cdot M(\widetilde k)\Big)
\]
and
\begin{eqnarray*}
\dot{\widetilde k}_j &=& \dot k_j -\partial_l  A_j(r)\,\dot r_l
\\&=& -\,\partial_{j}\phi(r) +
\partial_{l } E(\widetilde k)\,\partial_{j} A_l  (r) \\&&-\,
\epsi\, \partial_{k_l }\Big( \A (\widetilde k)\cdot F_{\rm Lor}(r,
\widetilde k) + M(\widetilde k)\cdot
B(r)\Big)\,\partial_{ j}A_l (r) \\
&& +\,\epsi \,\A_l (\widetilde k) \,\partial_{r_j} F_{{\rm Lor\,}
l }(r, \widetilde k) +\, \epsi\,\partial_j
  B (r)\cdot M (k-A(r)) -\partial_l  A_j(r)\,\dot r_l
\\&=&
-\,\partial_{j}\phi(r) + \dot r_l  \Big( \partial_j A_l (r) -
\partial_l  A_j(r)\Big) \\&&+\,\epsi \,\A_l (\widetilde
k)\, \partial_{r_j} F_{{\rm Lor\,} l }(r, \widetilde k) + \epsi\,
\partial_j B (r)\cdot M (\widetilde k)
\\
&=& -\,\partial_{j}\phi(r)  + \big( \dot r\vp B(r)\big)_j +\epsi
\,\A_l (\widetilde k)\, \partial_{r_j} F_{{\rm Lor\,} l }(r,
\widetilde k) + \epsi\, \partial_j B (r)\cdot M (\widetilde k)\,,
\end{eqnarray*}
which, in more compact form, read
\begin{eqnarray}
\dot r &=& \nabla E(\widetilde k) - \epsi \nabla_{\widetilde k}
\Big( \A (\widetilde k) \cdot F_{\rm Lor}(r, \widetilde k)
+  B(r)\cdot M(\widetilde k) \Big)\,,\nonumber\\ \label{53E1}\\
\dot{\widetilde k}&=& -\nabla \phi(r) + \dot r\vp B(r) +
\epsi\nabla_r \Big( \A (\widetilde k) \cdot F_{\rm
Lor}(r,\widetilde k)
+ B(r)\cdot M(\widetilde k) \Big)\,. \nonumber
\end{eqnarray}

As the next step we perform the change of coordinates induced by
$T$,
\be\label{53Qtrans}
q = r+ \epsi \A  (\widetilde k )\,,\qquad
 p= \widetilde k -A(r) + \epsi  \nabla_r \big( \A  (\widetilde k )\cdot   A(r)
 \big)\,,
\ee
and then switch to the kinetic momentum
\begin{eqnarray}\label{53Pitrans}
v & =& p - A(q) \nonumber \\&=& \widetilde k  + \epsi   \A_l
(\widetilde k)\nabla A_l  (r)   - \epsi   \A_l
(\widetilde k)\partial_l  A  (r)+\Or(\epsi^2)\nonumber\\
&=& \widetilde k + \epsi \,\A  (\widetilde k)\vp
B(r)+\Or(\epsi^2)\,,
\end{eqnarray}
where we used Taylor expansion.
The  inverse transformations are
\begin{eqnarray*}
r &= &q - \epsi \,\A  (v) +\Or(\epsi^2)\,,
\\
 \widetilde k& = &  v  - \epsi
\,\A  (v)\vp B(q) + \Or(\epsi^2)\,.
\end{eqnarray*}
Recall that we want to show that $(q,v)$ satisfy the semiclassical equations
of motion (\ref{Semi1}), where $q$ is identified with $r$ and $v$ with $\kappa$.
The new notation is  introduced here, only to make a clear distinction between 
the canonical variables $(r, k)$ in the reference representation and the canonical variables
$(q,p)$ in the original representation.

We  now substitute (\ref{53Qtrans}) and (\ref{53Pitrans}). In the
following computations we use several times Taylor expansion to
first order and drop terms of order $\epsi^2$. In particular in
the terms of order $\epsi$ one can replace $r$ by $q$ and
$\widetilde k$ by $v$. We find
\begin{eqnarray*}
\dot q_j &=& \dot r_j + \epsi \,\dot \A_j(v)\\
&=&  \partial_{j} E( v) -\epsi\,\Big(\A (v)\vp
B(q)\Big)_l  \partial_l \partial_j E(v) \\&& - \,\epsi\,
\partial_{v_j}\Big( \big(-\nabla\phi(q) + \nabla E(v)\vp
B(q)\big)_l  \A_l (v)
  + B(q)\cdot M(v)\Big)\\&& +\,\epsi\partial_l  \A_j
  \dot v_l \\&=&
 \partial_{j} E( v) - \epsi \dot v_l \Big( \partial_j \A_l -\partial_l  \A_j\Big) -\epsi\, B
 (q)\cdot
 \partial_j M (v) \\
 &=&
 \partial_{j} E( v) - \epsi \big(\dot v \vp \Omega(v)\big)_j   -\epsi\, B
 (q)\cdot
 \partial_j M  (v)\,,
\end{eqnarray*}
where it is used that $\dot v = F_{\rm Lor} + \Or(\epsi)$. Thus we
obtained the first equation of (\ref{Semi1}). For the second
equation we find
\begin{eqnarray*}
\dot v_j &= &\dot{\widetilde k}_j + \epsi \frac{\D}{\D
t}\,\Big(\A (v)\vp B(q)\Big) \\
&=&
-\,\partial_{j}\phi(q) + \epsi \A_l (v)\partial_l \partial_j \phi(q)
\\&&+\, \big( \dot q\vp B(q)\big)_j-
\epsi\big( \dot \A (v)\vp B(q)\big)_j - \epsi \Big(\dot q\vp \big(
\A_l (v)\partial_l  B(q)\big)\Big)_j
\\&&
+\,\epsi \,\A_l (v) \partial_{q_j} F_{{\rm Lor\,} l }(q, v) +
\epsi\,\partial_j B  (q)\cdot M (v)\\&&+\,
\epsi\big( \dot \A (v)\vp B(q)\big)_j+ \epsi \Big(\A (v)\vp \big( \dot
q_l \partial_l  B(q) \big)\Big)_j\\ &=&
-\,\partial_{j}\phi(q)+ \big( \dot q\vp B(q)\big)_j +
\epsi\, \partial_j B (q)\cdot M (v)\,,
\end{eqnarray*}
where the term
\[
\epsi \,\A(v) \Big( \partial_{q_j} F_{{\rm Lor\,} l }(q,
v)+\partial_l \partial_j \phi(q)\Big) = \epsi\, \A_l
(v) \Big( \dot q\vp \partial_j B(q)\Big)_l  + \Or(\epsi^2)
\]
cancels the remaining two terms. 
Changing back notation from $(q,v)$ to $(r,\kappa)$, this concludes the proof of
Theorem \ref{EgCor}.
 \end{proof}

\appendix
\section{Operator-valued Weyl calculus for $\tau$-equi\-va\-riant symbols}

The pseudodifferential calculus for scalar-valued symbols defined on 
the phase space $T^*\R^{d}=\R^{2d}$ can be translated 
to the phase space $T^*\T^d=\T^d\times \R^d$, $\T^d$ a flat torus,  
by restricting to periodic  functions and symbols. 
This
approach is used by G\'erard and Nier \cite{GeNi} in the context
of scattering theory in periodic media.

In this appendix we present a similar approach to Weyl
quantization of operator-valued symbols which are not exactly
periodic, but $\tau$-equivariant with respect to some nontrivial
representation $\tau$ of the group of lattice translations. We
obtain a pseudodifferential and semiclassical calculus which can
be applied to $\tau$-equivariant symbols like the Schr\"odinger
Hamiltonian with periodic potential in the Zak representation. In
particular, the full computational power of the usual Weyl
calculus is retained. The strategy is  to use the strong results
available for phase space $\R^{2d}$ by restricting to functions
which are $\tau$-equivariant in the configurational variable.

Let $\Gamma \subset \R^d$ be a regular lattice generated through
the basis $\{\gamma_1,\ldots,\gamma_d\}$, $\gamma_j\in\R^d$, i.e.\
\[
\Gamma =\Big\{ x\in\R^d: x= \textstyle{\sum_{j=1}^d}\alpha_j\,\gamma_j
\,\,\,\mbox{for some}\,\,\alpha \in \mathbb{Z}^d \Big\}\,.
\]
Clearly the translations on $\R^d$ by elements of $\Gamma$ form an
abelian group isomorphic to $\mathbb{Z}^d$. The centered
fundamental cell of $\Gamma$ is denoted as
\[
M=\Big\{ x\in\R^d: x= \textstyle{\sum_{j=1}^d}\alpha_j\,\gamma_j
\,\,\,\mbox{for}\,\,\alpha_j\in
[-\textstyle{\frac{1}{2},\frac{1}{2}}]
 \Big\}\,.
\]

Let $\Hi$ be a separable Hilbert space and let $\tau$ be a
representation of $\Gamma$ in $\B^{*}(\Hi)$, the group of
invertible elements  of $\B(\Hi)$ , i.e.\ a group homomorphism
\[
\tau: \Gamma \to \B^*(\Hi),\qquad  \gamma\mapsto \tau(\gamma)\,.
\]
 If more than one Hilbert space appears, then $\tau$ denotes a
collection of such representations, i.e.\ one on each Hilbert
space.\medskip

\noindent {\bf Warning:} In the application of the results of this
appendix to Bloch electrons the lattice $\Gamma$ corresponds to
the dual lattice $\Gamma^*$ in momentum space $\R^d$.
\medskip

Let $L_\gamma$ be the operator of translation by $\gamma\in\Gamma$
on $\Sch(\R^d,\Hi)$, i.e.\ $(L_{\gamma}\ph)(x)= \ph(x-\gamma)$,
and extend it by duality to  distributions, i.e.\ for
$T\in\Sch'(\R^d,\Hi)$ let $(L_\gamma T)(\ph)= T(L_{-\gamma}\ph)$.

\begin{definition}\label{ABequivdef}
A tempered distribution $T \in \Sch'(\R^d,\Hi)$ is said to be {\em
$\tau$-equivari\-ant} if
\[
L_\gamma T= \tau(\gamma) T \quad
\mbox{for all}\,\,\gamma\in\Gamma\,,
\]
where
$\big(\tau(\gamma)T\big)(\ph) = T\big(\tau(\gamma)^{-1} \ph\big)$
for $\ph\in\Sch(\R^d,\Hi)$. The subspace of $\tau$-equivari\-ant
distributions is denoted as $\Sch'_\tau$. Analogously we define
\[
\Hi_\tau = \Big\{ \psi\in L^2_{\rm loc}(\R^d,\Hi): \,
\psi(x-\gamma) = \tau(\gamma)\,\psi(x)\quad \mbox{for
all}\,\,\gamma\in\Gamma\Big\}\,,
\]
which, equipped with the
inner product
\[
\langle \ph,\psi\rangle_{\Hi_\tau} = \int_M \D x\, \langle
\ph(x),\psi(x)\rangle_{\Hi}\,,
\]
is a Hilbert space. Clearly
\[
C^\infty_\tau = \Big\{ \psi\in C^\infty(\R^d,\Hi): \,
\psi(x-\gamma) = \tau(\gamma)\,\psi(x)\quad \mbox{for
all}\,\,\gamma\in\Gamma\Big\}\,,
\]
is a dense subspace of $\Hi_\tau$.\er
\end{definition}

Notice that if $\tau$ is a unitary representation, then for any
$\ph, \psi \in \Hi_\tau$ the map $x\mapsto \langle
\ph(x),\psi(x)\rangle_{\Hi}$ is periodic, since
\[
 \langle  \ph(x-\gamma),\psi(x-\gamma)\rangle_{\Hi}
=  \langle \tau(\gamma) \ph(x), \tau(\gamma)\psi(x)\rangle_{\Hi} =
 \langle  \ph(x),\psi(x)\rangle_{\Hi}\,.
\]
Now that we have $\tau$-equivariant functions, we define
$\tau$-equivariant symbols. To this end we first recall the
definition of the standard symbol classes.

\begin{definition}
A function $w: \R^{2d} \to [0,+ \infty)$ is said to be an
\textbf{order function}, if there exist constants $C_0 > 0$ and
$N_0 > 0$ such that
\[
w(x) \leq C_0 \ \langle x-y \rangle^{N_0} \
w(y)
\]
 for every $x,y \in \R^{2d}$. \label{AADef order}\er
\end{definition}

It is obvious and will be used implicitly that the product of two
order functions is again an order function.

\begin{definition}
A function $A\in C^{\infty}(\R^{2d},\B(\Hi_1, \Hi_2))$ belongs to
the symbol class $S^w(\B(\Hi_1, \Hi_2))$ with order function $w$, if for every $\alpha
,\beta \in \N^{d}$ there exists a positive constant $C_{\alpha
,\beta }$ such that
\be\label{AASeminorms2}
\left\| (\partial_{q}^{\alpha
}\partial_{p}^{\beta}A)(q,p)\right\|_{\B(\Hi_1, \Hi_2)}\leq
C_{\alpha ,\beta }\ w(q,p)
\ee
for every $q,p \in \R^{d}$. \label{AADef Symb}\er
\end{definition}

\begin{definition} \label{AASemiSymbDef}
A map $A:[0,\epsi_{0})\rightarrow S_w(\B(\Hi_1,\Hi_2)),\epsi
\mapsto A_\epsi$ is a semiclassical symbol of
  order $w$, if there exists a sequence
$\{ A_j \}_{j\in \N}\subset A_j\in S^w(\B(\Hi_1,\Hi_2))$ such
that
\[
A \asymp \sum_{j=0}^\infty \epsi^j\,A_j\quad
\mbox{in}\quad S^w(\B(\Hi_1,\Hi_2))\,,
\]
which means that
for every $n\in \N$  and
for all $\alpha,\beta\in \N^d$ there exists a constant
$C_{\alpha ,\beta,n }$ such that for any $\epsi \in [0,\epsi_0)$
one has
\be \label{AASemisymb}
\Big\|\partial_{q}^{\alpha
}\partial_{p}^{\beta} \Big( A_\epsi(q,p) -\sum_{j=0}^{n-1}\,\epsi^j A_j(q,p) \Big) \Big\|_{\B(\Hi_1, \Hi_2)}
 \leq  \epsi^n \, C_{\alpha ,\beta,n }\ w(q,p)\,.
\ee
The space of semiclassical symbols of   order $w$
is denoted as $S^w(\epsi,\B(\Hi_1,\Hi_2) )$ or, if
clear from the context or if no specification is required, as  $S^w(\epsi)$.
The space of formal power series with coefficients in $S^w(\B(\Hi_1,\Hi_2))$ is
denoted as $M^w(\epsi,\B(\Hi_1,\Hi_2) )$.\er
\end{definition}

\begin{definition}
\label{ABDefequivsymbol}  A symbol $A_\epsi \in
S^w(\epsi,\B(\Hi_1,\Hi_2))$ is {\em $\tau$-equivariant} (more
precisely $(\tau_1, \tau_2)$-equivariant), if
\[
A_\epsi(q-\gamma,p) =
\tau_2(\gamma)\,A_\epsi(q,p)\,\tau_1(\gamma)^{-1} \quad \mbox{for
all}\,\,\gamma\in\Gamma\,.
\]
The space of $\tau$-equivariant symbols is denoted as $ S_{\tau}^w
(\epsi, \B(\Hi_1,\Hi_2))$.\er
\end{definition}

 Notice that  the coefficients in the asymptotic expansion of a
$\tau$-equivariant semiclassical symbol must be as well
$\tau$-equivariant, i.e.\ if $A_{\epsi}\asymp
\sum_{j=0}^\infty\epsi^j A_j$, $A_{\epsi}\in
S^w_{\tau}(\epsi,\B(\Hi_1,\Hi_2))$, then $A_j\in S_{\tau}^w(
\B(\Hi_1,\Hi_2))$.

Given any $\tau$-equivariant  symbol $A\in S_{\tau}^w (
\B(\Hi_1,\Hi_2))$, one can consider the usual Weyl quantization
$\widehat A$, regarded as an operator acting on
$\Sch'(\R^d,\Hi_1)$ with distributional integral kernel
\be
K_{A}(x,y)=\frac{1}{(2\pi \epsi)^{d}} \int_{\R^d}\D \xi\,
A\big({\textstyle \frac{1}{2}} (x+y),\xi \big)\ \E^{\I \xi \cdot
(x-y)/\epsi}\, . \label{ABKer2}
\ee
Notice that integral kernel associated to a
$\tau$-equivariant symbol $A$ is $\tau$-equivari\-ant in the
following sense:
\be\label{ABKperiodic}
K_A(x-\gamma,y-\gamma) =
\tau_2(\gamma)\,K_A(x,y)\,\tau_1(\gamma)^{-1} \quad \mbox{for
all}\,\,\gamma\in\Gamma\,.
\ee

The simple but important observation is that the space of
$\tau$-equivariant distributions is invariant under the action of
pseudodifferential operators with $\tau$-equivariant symbols.

\begin{proposition}\label{ABInvProp}
Let $A\in  S_{\tau}^w ( \B(\Hi_1,\Hi_2))$, then
\[
\widehat A\,
 \Sch'_{\tau_1}(\R^d,\Hi_1)
 \subset
\Sch'_{\tau_2}(\R^d,\Hi_2)\,.
\]
\end{proposition}
\begin{proof}
Since $\widehat A$ maps $\Sch'(\R^d,\Hi_1)$ continuously into
$\Sch'(\R^d,\Hi_2)$, we only need to show that $(L_\gamma \widehat
A T)(\ph) = (\tau_2(\gamma) \widehat A T)(\ph)$ for all $T\in
\Sch'_{\tau_1}(\R^d,\Hi_1)$ and $\ph\in\Sch(\R^d,\Hi_2)$.

To this end notice that as acting on $\Sch(\R^d,\Hi_2)$ one finds
by direct computation using (\ref{ABKer2}) that
$\widehat{A^*}\,L_\gamma=L_\gamma
\,(\tau_1(\gamma)^{-1})^*\,\widehat{
  A^*}\,\tau_2(\gamma)^*$. Indeed, let $\psi\in \Sch(\R^d,\Hi_2)$,
then
\begin{eqnarray*}
\big(\widehat{ A^*}\,L_\gamma\,\psi\big)(x) &=& \int_{\R^d} \D y\,
K_{A^*} (x,y) \,\psi(y-\gamma) = \int_{\R^d} \D y\, K_{A^*}
(x,y+\gamma)
\,\psi(y) \\
&=&  \int_{\R^d} \D y\, (\tau_1(\gamma)^{-1})^*\, K_{A^*}
(x-\gamma,y)\, \tau_2(\gamma)^*
\,\psi(y)\\
&=& \big( L_\gamma\, (\tau_1(\gamma)^{-1})^*\,\widehat{ A^*}\,
\tau_2(\gamma)^*\,\psi\Big)(x)
\end{eqnarray*}
Hence, using the fact that $\tau$ is a representation and that
$L_\gamma T=\tau_1(\gamma)T$,
\begin{eqnarray*}
(L_\gamma \widehat A T)(\ph) &=& T(\widehat A^*\,L_{-\gamma}\,\ph)
= T( L_{-\gamma}\,\tau_1(\gamma)^*\,\widehat
A^*\,(\tau_2(\gamma)^{-1})^*\,\ph)
\\& =& (\tau_2(\gamma)\,\widehat A\,\tau_1(\gamma)^{-1}\,L_\gamma
\,T)(\ph)=
 (\tau_2(\gamma)\,\widehat A\,T)(\ph)\,.
\end{eqnarray*}
   \end{proof} 

For the convenience of the reader we also recall the definition
and the basic result about the Weyl product of semiclassical
symbols. For a proof see e.g.\ \cite{DiSj}.

\begin{proposition}\label{AAPropMoyalproduct}
Let $A\in S^{w_1} (\epsi, \B(\Hi_2,\Hi_3))$ and $B\in S^{w_2}
(\epsi,\B(\Hi_1,\Hi_2))$, then $\widehat A\widehat B = \widehat C$, with
$C\in S^{w_1 w_2}(\epsi ,\B(\Hi_1,\Hi_3))$ given through
\be \label{AACDEF}
C(\epsi,q,p) = \exp\left( \frac{\I\,\epsi}{2}\,(\nabla_p\cdot \nabla_x -
\nabla_\xi\cdot\nabla_q )\right)
A(\epsi,q,p)B(\epsi,x,\xi)\Big|_{x=q,\xi=p} \hspace{-1mm}=: A\,\widetilde\sharp  \,B\,.
\ee
\end{proposition}

The corresponding product on the level of the formal power series
is called Moyal product and denoted as
\[
\sharp  :M^{w_1} (\epsi, \B(\Hi_2,\Hi_3))\times
M^{w_2}(\epsi,\B(\Hi_1,\Hi_2)) \to M^{w_1 w_2}(\epsi
,\B(\Hi_1,\Hi_3))\,.
\]
The $\tau$-equivariance of symbols is preserved
under the pointwise product, the Weyl product and the Moyal
product.

\begin{proposition}\label{ABProSymbComp}
Let $A_\epsi \in  S^{w_1}_{\tau} (\epsi, \B(\Hi_2,\Hi_3))$ and
$B_\epsi \in S^{w_2}_{\tau} (\epsi, \B(\Hi_1,\Hi_2))$, then
$A_\epsi B_\epsi\in S^{w_1 w_2}_{\tau} (\epsi , \B(\Hi_1,\Hi_3))$
and $A_\epsi\,\widetilde{\sharp  }\,B_\epsi \in S^{w_1 w_2}_{\tau}
(\epsi, \B(\Hi_1,\Hi_3))$.
\end{proposition}

\begin{proof}
One has
\begin{eqnarray*}
A_\epsi(q-\gamma,p)B_\epsi(q-\gamma,p) &=&\tau_3(\gamma)
A_\epsi(q,p)\tau_2(\gamma)^{-1}\tau_2(\gamma)
B_\epsi(q,p)\tau_1(\gamma)^{-1}
\\&=&
\tau_3(\gamma) A_\epsi(q,p) B_\epsi(q,p)\tau_1(\gamma)^{-1}\,,
\end{eqnarray*}
which shows  $A_\epsi B_\epsi\in  S^{w_1 w_2}_{\tau} (\epsi,
\B(\Hi_1,\Hi_3))$ and inserted into (\ref{AACDEF}) yields
immediately also $A_\epsi\,\widetilde\sharp  \,B_\epsi \in S^{w_1
w_2}_{\tau} (\epsi, \B(\Hi_1,\Hi_3))$.
  \end{proof} 
  An analogous
statement holds for the Moyal product of formal symbols.

A not completely obvious fact is the following variant of the
Calderon-Vaillan\-court theorem.

\begin{theorem} \label{ABThCaldVailltorus}
Let $A\in S^{1}_{\tau}(\B(\Hi))$ and $\tau_1, \tau_2$   unitary
representations of $\Gamma$ in $\B(\Hi)$, then $\widehat A\in
\B(\Hi_{\tau_1},\Hi_{\tau_2})$ and for $A_\epsi \in
S^{1}_{\tau}(\epsi, \B(\Hi))$ we have that
\[
\sup_{\epsi\in[0,\epsi_0)} \|\widehat
A_\epsi\|_{\B(\Hi_{\tau_1},\Hi_{\tau_2} )}<\infty\,.
\]
\end{theorem}
\begin{proof}
Fix $n>d/2$ and let $w(x) = \langle x\rangle^{-n}$. We consider
the weighted $L^2$-space
\[
L^2_w = \left\{ \psi\in L^2_{\rm loc}(\R^d,\Hi):\,\,\int_{\R^d}\D
x\, w(x)^2 |\psi(x)|^2 <\infty\right\}\,.
\]
Let $j=1,2$, then $\Hi_{\tau_j}\subset  L^2_w$ and for any
$\psi\in \Hi_{\tau_j}$ one has the norm equivalence
\be\label{ABNE}
C_1\, \|\psi\|_{\Hi_{\tau_j}} \leq \|\psi\|_{L^2_w} \leq C_2 \,
\|\psi\|_{\Hi_{\tau_j}}
\ee
for appropriate constants $0<C_1,C_2<\infty$. The first inequality
in (\ref{ABNE}) is obvious and the second one follows by
exploiting $\tau_j$-equivariance of $\psi$ and unitarity of
$\tau_j$:
\begin{eqnarray*}
\|\psi\|_{L^2_w}^2 &=& \sum_{\gamma\in\Gamma}\int_{M+\gamma}\D x\,
w(x)^2\,  \|\tau_j(\gamma)^{-1}\psi(x)\|^2_\Hi =
\sum_{\gamma\in\Gamma}\int_{M+\gamma}\D x\,
w(x)^2 \, \|\psi(x)\|^2_\Hi\\
&\leq& \sum_{\gamma\in\Gamma} \sup_{x\in M+\gamma}\left\{
w(x)^2\right\}\,\int_M\D x\,\|\psi(x)\|^2_\Hi\leq
C_2\,\|\psi\|_{\Hi_{\tau_j}}\,.
\end{eqnarray*}
According to (\ref{ABNE}) it suffices to show that $\widehat
A\in\B(L^2_w)$ and to estimate the norm of $\widehat A_\epsi$ in
this space.

Let $\psi\in C^\infty_{\tau_1}(\R^d,\Hi)$, then by the general
theory $\widehat A \psi$ is smooth as well (see \  \cite{Fo},
Corollary 2.62) and thus, according to Proposition
\ref{ABInvProp}, $\widehat A \psi\in C^\infty_{\tau_2}(\R^d,\Hi)$.
Hence we can use (\ref{ABNE}) and find
\[
\big\| \widehat A\psi \big\|_{L^2_w} = \big\| w \widehat A\psi
\big\|_{L^2} \leq
 \big\| w \widehat A w^{-1} \big\|_{\B(L^2)} \big\|w\psi\big\|_{L^2} =
\big\| w \widehat A w^{-1} \big\|_{\B(L^2)}
\big\|\psi\big\|_{L^2_w}\,.
\]
However,  by Proposition \ref{AAPropMoyalproduct}, we have that
$w\,\widetilde\sharp  \,A\,\widetilde\sharp  \,w^{-1}\in S^{1}(\epsi,
\B(\Hi))$. Thus from the usual Calderon-Vaillancourt theorem it
follows that
\[
 \big\| w \ \widehat A \ w^{-1} \big\|_{\B(L^2)} \leq C_d\, \big\|
\, w\,\widetilde\sharp  \,A\,\widetilde\sharp  \,w^{-1} \big\|_{C^{2d+1}_{\rm
b}(\R^{2d})}\,.
\]
This shows that for $A\in S_{\tau}^{1}(\B(\Hi))$ we have
$\widehat A\in \B(\Hi_{\tau_1},\Hi_{\tau_2})$. With
$w\,\widetilde\sharp  \,A_\epsi\,\widetilde\sharp  \,w^{-1} \in  S^{1}(\epsi,
\B(\Hi))$ for $A_\epsi \in S^{1}_{\tau}(\epsi,\B(\Hi))$, we
conclude that
\[
\sup_{\epsi\in[0,\epsi_0)} \|\widehat
A_\epsi\|_{\B(\Hi_{\tau_1},\Hi_{\tau_2})}<\infty
\]
by the same argument.
 \end{proof}

\begin{remark} It is clear from the proof that   the previous
result still holds true under the weaker assumption that $\tau_1$
and $\tau_2$ are uniformly bounded, i.e. that
\[ \sup_{\gamma \in \Gamma} \|
\tau_j(\gamma) \|_{\B(\Hi)} \leq C\,,\quad j=1,2\,. 
\] \er
\end{remark}

Finally we would also like to show that for $A\in S_{\tau}^1
(\B(\Hi))$ the adjoint of $\widehat A$ as an operator in
$\B(\Hi_\tau)$, denoted by $\widehat A^\dagger$, is  given through
the quantization of the pointwise adjoint, i.e.\ through
$\widehat{A^*}$. Here it is crucial that $\tau$ is a unitary
representation.

\begin{proposition}
Let  $S_{\tau}^1( \B(\Hi))$ with a unitary representation $\tau$
(with $\tau_1=\tau_2=\tau$) and let
 $\widehat A^\dagger$ be the adjoint of $\widehat A\in\B(\Hi_\tau)$,
 then $\widehat A^\dagger = \widehat{A^*}$.
\end{proposition}
\begin{proof}
Let $\psi\in \Hi_\tau$ and $\ph \in C^\infty_\tau$ such that
$\widetilde \ph := {\bf 1}_M\,\ph \in C^\infty_0(\R^d,\Hi)$, where
${\bf 1}_M$ denotes the characteristic function of the set $M$.
Such $\ph$ are dense in $\Hi_\tau$ and the corresponding
$\widetilde \ph$ can be used as a test function:
\begin{eqnarray*}
\big\langle\ph, \widehat A\psi\big\rangle_{\Hi_\tau} &=& \int_M\D
x\,\big\langle \ph(x),\, (\widehat A\psi)(x)\big\rangle_{\Hi} =
\int_{\R^d} \D x\, \big\langle \widetilde \ph(x),\,  (\widehat
A\psi)(x)\big\rangle_{\Hi}\\
&=& \int_{\R^d} \D x\, \big\langle (\widehat{A^*} \widetilde
\ph)(x),\,
\psi(x)\big\rangle_{\Hi}\\
&=& \int_{\R^d} \D x\, \Big\langle \int_{\R^d}\D y\, K_{A^*}(x,y)
\widetilde
\ph(y),\,\psi(x)\Big\rangle_{\Hi}\\
&=& \int_{\R^d} \D x\, \Big\langle \int_M\D y\, K_A^*(x,y)
\widetilde
\ph(y),\,\psi(x)\Big\rangle_{\Hi}\\
&=& \int_M \D x\,\sum_{\gamma\in\Gamma}
 \Big\langle \int_M\D y\, K_A^*(x+\gamma,y) \widetilde
\ph(y),\,\psi(x+\gamma)\Big\rangle_{\Hi}\\
&=& \int_M \hspace{-1pt}\D x\,\sum_{\gamma\in\Gamma}
 \Big\langle \hspace{-1pt}\int_M\hspace{-1pt}\D y\,\tau^{-1}\hspace{-1mm}(\gamma)
K_A^*(x,y-\gamma)\tau(\gamma)
\widetilde \ph(y),\,\tau^{-1}(\gamma)\psi(x)\Big\rangle_{\Hi}\\
&=& \int_M \D x\,\sum_{\gamma\in\Gamma}
 \Big\langle \int_M\D y\, K_A^*(x,y-\gamma)
 \ph(y-\gamma),\,\psi(x)\Big\rangle_{\Hi}\\
&=& \int_M \D x\,
 \Big\langle \int_{\R^d}\D y\, K_A^*(x,y)
 \ph(y),\,\psi(x)\Big\rangle_{\Hi}\\
&=& \int_M \D x\,
 \big\langle (\widehat{A^*}\ph)(x),\,\psi(x)\big\rangle_{\Hi}
= \big\langle\widehat{ A^*}\ph, \psi\big\rangle_{\Hi_\tau}
\end{eqnarray*}
In particular, we used the $\tau$-equivariance of the kernel
(\ref{ABKperiodic}) and of the functions in $\Hi_\tau$ and the
unitarity of $\tau$. By density we have $\widehat{A^*} = \widehat
A^\dagger$.
  \end{proof}


\section{Hamiltonian formulation for the refined semiclassical model}

The dynamical equations (\ref{Semi1}), which define the
$\epsi$-corrected semiclassical model, can be written as
\begin{equation} \label{SC model}
\begin{array}{ccc}
    \dot r & = & \nabla_\kappa H_{\rm sc}(r,\kappa)  -   \epsi \, \dot \kappa \times \Omega_n(\kappa)\,, \\
    &&\\
    \dot \kappa & = & \hspace{-3mm}-  \nabla_r H_{\rm sc}(r,\kappa) +   \dot r \times B(r)
\end{array}
\end{equation}
with
\[ \label{SC energy}
H_{\rm sc}(r,\kappa) := E_n(\kappa) + \phi(r) - \epsi \, M_n(\kappa) \cdot
B(r)\,.\]
Recall that we are using the notation introduced in Remark
\ref{NotRem} and that $B$ and $\Omega_n$ are
 the 2-forms corresponding  to the magnetic field and
to the curvature of the Berry connection, i.e.\ in components
\[
B(r)_{ij} = \big( \partial_{i} A_j - \partial_j A_i \big)(r)
\]
for $i,j \indexd$, and
\[
\Omega_n(\kappa)_{ij} = \big( \partial_{i} \mathcal{A}_j - \partial_j
\mathcal{A}_i \big)(\kappa)\,.
\]
We fix the system of coordinates $z=(r,\kappa)$ in $\R^{2d}$. The
standard symplectic form $\Theta_0 = \Theta_{0}(z)_{lm}\, \D z_m
\wedge \D z_l $, where $l,m \in\{ 1,\ldots ,2d\}$, has
coefficients given by the constant matrix
\[ \label{Symplectic standard} \Theta_{0}(z)= \left(
\begin{array}{cc}
  0 & -\mathbb{I} \\
  \mathbb{I} & 0
\end{array}\right)\,,
\]
where $\mathbb{I}$ is the identity matrix in $\mathrm{Mat}(d,\R)$.
The symplectic form, which turns (\ref{SC model}) into Hamilton's
equation of motion for $H_{\rm sc}$, is given by the  2-form
$\Theta_{B,\, \epsi}= \Theta_{B,\,\epsi}(z)_{lm}\,  \D z_m \wedge
\D z_l$ with coefficients \be \label{Symplectic matrix}
\Theta_{B,\, \epsi}(r,\kappa)= \left(
\begin{array}{cc}
  B(r) & -\mathbb{I} \\
  \mathbb{I} &  \epsi \ \Omega_n(\kappa)
\end{array}\right)\,.
\ee For $\epsi=0$ the 2-form $\Theta_{B,\, \epsi}$ coincides with
the magnetic symplectic form $\Theta_{B}$ usually employed to
describe in a gauge-invariant way  the motion of a particle in a
magnetic field (\cite{MaRa}, Section~6.6). For $\epsi$ small
enough, the matrix (\ref{Symplectic matrix})   defines a
symplectic form, i.e.\ a closed non-degenerate 2-form. Indeed,
since $\det \Theta_{B} = 1$ it follows that, for $\epsi$ small
enough, $\Theta_{B,\, \epsi}$ is   not degenerate. In particular
it is sufficient to choose
\[
\epsi < \sup_{r,\kappa \in \R^d} \big( \|  B(r) \, \Omega_n(\kappa) \| + \|
\Omega_n(\kappa) \| \big)\,.
\] 
The closedness of $\Theta_{B,\, \epsi}$ follows from the fact that
$B$ and $\Omega_n$ correspond to closed 2-forms over $\R^d$.

 With these definitions the corresponding Hamiltonian equations are
\[ \Theta_{B,\, \epsi}(z) \ \dot{z} = \D H_{\rm sc}(z)\,,
\]
or equivalently
\[
\left(
\begin{array}{cc}
  B(r) & -\mathbb{I} \\
  \mathbb{I} & \epsi \ \Omega_n(\kappa)
\end{array}\right)
\left(
\begin{array}{c}
  \dot r \\
  \dot \kappa
\end{array}\right) =
\left(
\begin{array}{c}
  \nabla_r H(r,\kappa) \\
  \nabla_\kappa H(r,\kappa)
\end{array}\right)\,,
\]
which agrees with (\ref{SC model}). We notice that this discussion
remains valid if  $\Omega_n$ admits a potential only locally, as
it happens generically for magnetic Bloch bands.


\noindent {\bf Acknowledgements.} \,
 G.\ P. is grateful for financial
support by the Research Training Network HYKE of the European
Union and by the  Priority Program  ``Analysis, Modeling and
Simulation of Multiscale Problems'' of the Deutsche
Forschungsgemeinschaft.

\end{document}